\documentclass[10pt,letterpaper]{article}
\usepackage[top=0.85in,left=2.75in,footskip=0.75in]{geometry}
\usepackage{changepage}
\usepackage[utf8]{inputenc}
\usepackage{textcomp,marvosym}
\usepackage{fixltx2e}
\usepackage{amsmath,amssymb}
\usepackage{cite} 
\usepackage{nameref,hyperref}
\usepackage{microtype}
\DisableLigatures[f]{encoding = *, family = * }
\usepackage{rotating}
\raggedright
\usepackage[aboveskip=1pt,labelfont=bf,labelsep=period,justification=raggedright,singlelinecheck=off]{caption}
\makeatletter
\renewcommand{\@biblabel}[1]{\quad#1.}
\makeatother
\date{}
\usepackage{lastpage,fancyhdr,graphicx}
\usepackage{epstopdf}
\fancyheadoffset[L]{2.25in}
\fancyfootoffset[L]{2.25in}
\usepackage{graphics}             
\newcommand{\apj}{ApJ}            
\newcommand{\apjl}{ApJ}           
\newcommand{\mnras}{MNRAS}        
\newcommand{\val}{\it}            
\newcommand{\aste}{^{\mathrm{o}}} 
\newcommand{\sivu}{\clearpage}    
\begin{document}
\vspace*{0.35in}
\begin{flushleft}
{\Large
\textbf\newline{Shifting Milestones of Natural Sciences: \\
The Ancient Egyptian Discovery of Algol's Period Confirmed }}
\newline
\\
Lauri Jetsu\textsuperscript{1,*},
Sebastian Porceddu\textsuperscript{1}
\\
\bigskip
\bf{1} Department of Physics, P.O. Box 64, 
       FI-00014 University of Helsinki, Finland 
\\
*E-mail: lauri.jetsu@helsinki.fi

\end{flushleft}
\section*{Abstract}
The Ancient Egyptians wrote Calendars of Lucky and Unlucky 
Days that assigned  astro\-nom\-ically                      
influenced prognoses for each day of the year. 
The best preserved of these calendars
is the Cairo Calendar (hereafter CC)            
dated to 1244--1163 B.C.
We have presented evidence that the 2.85 days 
period in the lucky prognoses 
of CC is equal to that of the eclipsing binary Algol 
during this historical era.                           
We wanted to find out the vocabulary that represents Algol in 
the mythological texts of CC. 
Here we show that Algol was represented as Horus
and thus signified both divinity and kingship. 
The texts describing the actions of Horus are 
consistent with the course of events witnessed 
by any naked eye observer of Algol.
These descriptions support our claim                
that CC is the oldest preserved 
historical document of the discovery of a variable star. 
The period of the Moon, 29.6 days, 
has also been discovered in CC.                  
We show that the actions of Seth were connected to this period, 
which also strongly regulated the times described as lucky 
for Heaven and for Earth. 
Now, for the first time, 
periodicity is discovered in the descriptions of the days in CC. 
Unlike many previous attempts                  
to uncover the reasoning behind the myths of individual days, 
we discover the actual 
rules in the appearance and behaviour of deities during the whole year. 


\section*{Introduction}

The Ancient Egyptians referred to celestial events 
indirectly\cite{Lei89,Wel94,Cla95,Smi12} by relating 
them to mythological events.
Many prognoses in the Calendars of Lucky and Unlucky Days 
have been connected to astronomical observations\cite{Lei89,Wel01,Kra02,Har02}.
Such connections between astronomical events and prognosis texts
have been uncovered in most cases
only for individual days\cite{Bru70,Tro89,Kra02}.
The $P_{\mathrm{M}}=29.6$ days period of the 
{\val Moon} has been discovered in CC\cite{Por08}. 
We have claimed that this document also contains the 
$P_{\mathrm{A}}=2.85$ days period of
the eclipsing binary {\val Algol}\cite{Jet13}.
However, it not a straightforward task to identify
those indirect mythological references that
are influenced by {\val Algol} in CC.
Here we present a statistical analysis that reveals which CC prognosis texts
describe {\val Algol}'s regular variability.

The Ancient Egyptian year contained 12 months ($M$) of 30 days ($D$)
and five additional ``epagomenal'' days.
CC gives three prognoses for each $D$ of every $M$
(G = ``gut''= ``good'' and S = ``schlecht'' = ``bad'')\cite{Lei94,Jet13}.
CC also gives textual descriptions of the daily prognoses
(S1 Fig).
We study the dates of 28 selected words (hereafter SWs)
in these mythological texts of CC.
The dates are transformed into series of
time points $t_{\mathrm{i}}$ with equation (\ref{timetwo}).
The $P_{\mathrm{A}}$ and $P_{\mathrm{M}}$ signals were originally discovered\cite{Jet13}
from six large samples of lucky prognoses ($n=6\times564=3384$).
We use these six samples to determine the zero epochs  $t_{\mathrm{E}}$ 
of equation (\ref{epoch}) 
for the $P_{\mathrm{A}}$ and $P_{\mathrm{M}}$ signals. 
The time points leading to the discovery of these signals 
were close to phase, $\phi=0$, of equation (\ref{phase}) 
using the ephemerides of equations (\ref{aephe}) and (\ref{mephe})
based on these zero epochs $t_{\mathrm{E}}$.
The lucky prognoses of each SW are a subsample of
the above mentioned large samples of lucky prognoses. 
We compute an impact parameter $z_{\mathrm{x}}$ for 
the $t_{\mathrm{i}}$ of each SW with equation (\ref{zx}).
The time points $t_{\mathrm{i}}$ of the lucky prognoses of any particular SW may strengthen 
(if $z_{\mathrm{x}}>0$)
or weaken 
(if $z_{\mathrm{x}}<0$) the $P_{\mathrm{A}}$ and $P_{\mathrm{M}}$ signals.
The impact parameter $z_{\mathrm{x}}$ is used for identifying
the SWs having lucky prognoses close to
phase, $\phi=0$, computed with 
the ephemerides of equations (\ref{aephe}) and (\ref{mephe}).
We will show that {\val Algol} and the {\val Moon} 
were at their brightest  
close to phase $\phi = 0$ with these two ephemerides.
Hence, {\val Algol}'s eclipse and the New {\val Moon} occurred 
close to $\phi=0.5$.

Our statistical analysis also confirms two general things 
regarding the origin of the mythological texts of CC. 
First, the appearances and feasts of various deities are 
not independent of the prognoses, or randomly assigned, 
but regulated by the same periodic patterns. 
Second, the deities are used to represent 
the same astronomical phenomena that were also used to 
choose the prognoses for the days of the year.

\section*{Materials}

In this section, we transform the dates of 28 SWs
in the mythological texts of CC into series of time points $t_{\mathrm{i}}$.
Our main aim is that all stages of the production of these data 
can be replicated.
With these instructions, similar series of time points can be produced 
for any particular SW in CC or other similar calendars, 
where the SW dates are available.
We create the data in two stages: 
\nameref{identify} 
and
\nameref{transform}.

\subsection*{Identification of SW dates}
\label{identify}

CC is the best preserved Calendar of Lucky and unlucky Days.
As in our two previous studies\cite{Por08,Jet13},
we use the best preserved continuous calendar found on
pages recto III-XXX and verso I-IX of papyrus Cairo 86637.
There are two CC translations, 
in English by Bakir\cite{Bak66} and in German by Leitz\cite{Lei94}.
Our SWs have been identified according to the 
hieroglyphic transcription in Leitz\cite{Lei94}
 and the two aforementioned translations. 
In case
of discrepancy we have consulted the photocopies 
of the original hieratic text given by Leitz\cite{Lei94}. 
For the sake of convenience, we quote sentences according 
to Bakir's English translation despite its imperfections
 because there is neither space nor reason 
to discuss the linguistic details of the text in the present article. 
This approach should ascertain that our study of the CC 
sentences is objective.
In other words, we do not ourself translate any CC sentences into English,
but we do check which individual Ancient Egyptian SWs were also 
identified by Bakir\cite{Bak66} and Leitz\cite{Lei94}.
There is only one exception to our sentence quotation rule, i.e.
the CC text connected to {\val Horus} where
Bakir\cite{Bak66} did not identify {\val Horus},
but Leitz\cite{Lei94} and we did
(\nameref{Alucky}: the text at date $g_{\mathrm{i}}(1,10)$).

Naturally, we can not analyse all words in CC.
Our main selection criterion is to include deities,
nouns or locations that could have been used 
to indirectly describe
 periodic phenomena, 
due to their significant mythological properties 
and multiple occurrences in the text. 
Our list of SWs is not absolute 
and we give all the necessary information 
for other researchers to repeat our experiment 
on other SWs we may have ignored.
Our 28 SWs in Ancient Egyptian language are given in Table 1. \\

\clearpage

{\bf Table 1. List of SWs in Ancient Egyptian language.}
\begin{figure}[h]
\resizebox{12.5cm}{!}{\includegraphics{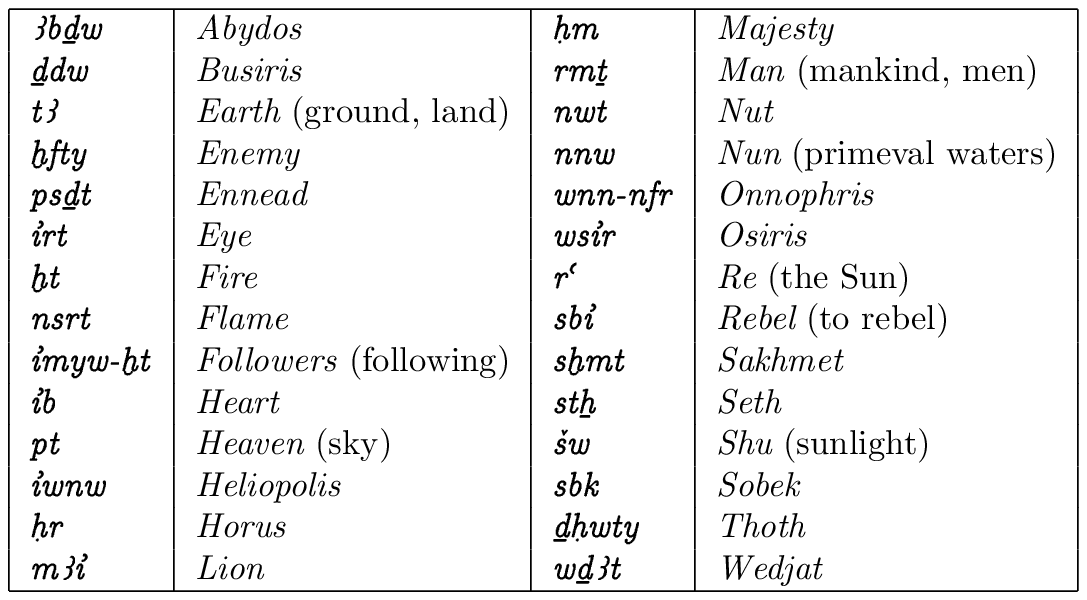}} 
\end{figure}

\setcounter{table}{1}

\clearpage

\noindent
We do not use the occurrences of our SWs in compound words and 
composite deities (e.g. House of Horus or Ra-Horakhti),
because it is uncertain to which word,
if not both, the prognosis is connected to.
Our identifications of 28 SWs in CC
are given in Table \ref{table2}.
It shows that all our 460 SW date identifications are the same 
as those made by Leitz\cite{Lei94}
(Column 5: $460 \times\!$ ``Yes'').
However, 21 of our identifications were not made by 
Bakir\cite{Bak66}
(Column 6: $21 \times\!$ ``No'': 
$1 \times$ ``Earth'', 
$2 \times$ ``Enemy'',
$4 \times$ ``Fire'',
$12 \times$ ``Heart'', 
$1 \times$  ``Horus'' 
and  
$1 \times$ ``Osiris'').
Fortunately, most days have combinations ``GGG'' or ``SSS''
and we know that the lucky or unlucky SW prognosis is certainly correct.
We ignore the heterogeneous
combinations ``HET'' (like ``SSG''at $D=6$ and $M=1$), 
because the correct SW prognosis is uncertain.
The dates with an unknown prognosis combination, ``-~-~-'', are naturally also ignored.
Our notations for the number of lucky and unlucky dates for each SW 
are $n_{\mathrm{G}}$ and $n_{\mathrm{S}}$.
For example, ``Abydos'' has $n_{\mathrm{G}}=3$ and $n_{\mathrm{S}}=2$.

\begin{table}[!ht]
\caption{
{\bf SWs identified in CC.}}
\begin{tabular}{|l|r|r|c|c|c|}
\hline
SW & $D$ & $M$ & Prog & Ltz & Bkr \\
\hline
Abydos  &   13  & 3 & SSS & Yes & Yes \\
\hline
Abydos  &   17  & 3 & -~-~- & Yes & Yes \\
\hline
Abydos  &   11  & 4 & GGG & Yes & Yes \\
\hline
Abydos  &   18  & 5 & GGG & Yes & Yes \\
\hline
Abydos  &   27  & 6 & -~-~- & Yes & Yes \\
\hline
Abydos  &   28  & 7 & GGG & Yes & Yes \\
\hline
Abydos  &   13  & 8 & SSS & Yes & Yes \\
\hline
Abydos  &   23  & 8 & -~-~- & Yes & Yes \\
\hline
Busiris &   26  & 2 & SSS & Yes & Yes \\
\hline
Busiris &   14  & 5 & SSS & Yes & Yes \\
\hline
Busiris &  26   & 5 & SSS & Yes & Yes \\
\hline

\end{tabular}
\begin{flushleft}
The selected word (SW) identified on day $(D)$ of month $(M)$ in CC. 
The daily prognosis combinations (Prog)
are ``GGG'' (All lucky), ``SSS'' (All unlucky),
``-~-~-'' (All unknown) or ``HET'' (Heterogeneous).
The same SW was identified at the same date by 
Leitz\cite{Lei94}
(Ltz=''Yes'' or ``No'') 
and by 
Bakir\cite{Bak66}
(Bkr=''Yes'' or ``No'').
The twelve first lines of all 460 lines are shown here for guidance 
regarding the contents of this ASCII file which
can be downloaded on Dryad (http://dx.doi.org/10.5061/dryad.tj4qg).
\end{flushleft}
\label{table2}
\end{table}

\sivu

\subsection*{Transformation of SW dates 
into series of time points}
\label{transform}

The dating of CC does not influence
the results of our currect analysis, because
we transform the time points to unit vectors with equation (\ref{unit}).
The mutual directions between these unit vectors
do not depend on the chosen zero epoch $t_0$ in time.
Adding any positive or negative constant value to these time
points rotates all unit vectors with the same constant angle.
Hence, our significance estimates of equations (\ref{qz})
and (\ref{qb}) do not depend on the connection
between Gregorian and Egyptian days.
The only assumption made in our equation (\ref{timetwo}) below is that 
the separation between two subsequent days is exactly one day
during the particular year that CC happens to describe.

The transformation relations in 
equations (2) and (3) of Jetsu et al.\cite{Jet13} were
\begin{eqnarray}
t_{\mathrm{i}} = N_{\mathrm{E}}-1+a_{\mathrm{i}}, 
\label{timeone}
\end{eqnarray}
\noindent
where $N_{\mathrm{E}}= 30 (M-1)+D$ and $a_{\mathrm{i}}$ was a decimal part. 
This decimal part  $a_{\mathrm{i}}$ was different for each of 
the three parts of the day.
The $a_{\mathrm{i}}$ values depended on the chosen transformation 
between Egyptian and Gregorian year, and on the chosen day division.
The $P_{\mathrm{A}}$ and $P_{\mathrm{M}}$ signals were 
discovered in samples of series of time points
SSTP=1, 3, 5, 7, 9 and 11 in Jetsu et al.\cite{Jet13}.
The size of each sample was $n=564$.
The period analysis results were the same for all these six samples,
although their $a_{\mathrm{i}}$ values were different for every $N_{\mathrm{E}}$.
The time points $t_{\mathrm{i}}$ of these six samples are given 
in Table \ref{table3}.

\begin{table}[!ht]
\caption{
{\bf The time points $t_{\mathrm{i}}$ of lucky prognoses in Jetsu et al.\cite{Jet13}.}}
\begin{tabular}{|c|c|c|c|c|c|}
\hline
SSTP=1    & SSTP=3 & SSTP=5 & SSTP=7 & SSTP=9 &SSTP=11 \\ 
\hline
    0.080 &  0.095 &  0.076 &  0.120 &  0.142 & 0.114 \\
\hline
    0.239 &  0.284 &  0.227 &  0.359 &  0.426 & 0.341 \\
\hline
    0.399 &  0.473 &  0.379 &  0.739 &  0.784 & 0.727 \\
\hline
    1.080 &  1.095 &  1.076 &  1.120 &  1.142 & 1.113 \\
\hline
    1.240 &  1.284 &  1.227 &  1.360 &  1.425 & 1.340 \\
\hline
\end{tabular}
\begin{flushleft} The $t_{\mathrm{i}}$ values of 
SSTP=1, 3, 5, 7, 9 and 11 from Table 3 in Jetsu et al.\cite{Jet13}.
The five first lines of all 534 lines are shown here for guidance 
regarding the contents of this ASCII file which
can be downloaded on Dryad (http://dx.doi.org/10.5061/dryad.tj4qg).
\end{flushleft}
\label{table3}
\end{table}

\sivu

\noindent
The mean of the decimal parts $a_{\mathrm{i}}$ 
of all these $n= 6 \times 564 = 3384$ 
values of  $t_{\mathrm{i}}$ is $m_{\mathrm{t}}=0.33$.
In this study, the time point for an SW at the day $D$ 
of the month $M$ in CC is therefore computed from
\begin{eqnarray}
t_{\mathrm{i}}=t_{\mathrm{i}}(D,M)=N_{\mathrm{E}}-1+m_{\mathrm{t}}.
\label{timetwo}
\end{eqnarray}
This accuracy is sufficient, 
because we do not know to which part or parts
of the day each SW refers to
($\sigma_{\mathrm{t}} \approx \pm 0.^{\mathrm{d}}5$) and
some prognosis texts may refer to the previous or the next day 
($\sigma_{\mathrm{t}} \approx \pm 1.^{\mathrm{d}}5$).
The  $t_{\mathrm{i}}$ of Table \ref{table3} $(n=6 \times 564 =3384)$ 
are also later used to determine
the zero epochs $t_{\mathrm{E}}$ for the ephemerides connected to 
the $P_{\mathrm{A}}$ and $P_{\mathrm{M}}$ signals
(equations (\ref{aephe}) and (\ref{mephe})).
Our ``synchronization'' of time points of
equations (\ref{timeone}) and (\ref{timetwo})
ensures that these ephemerides enable us to identify the SWs
connected to the $P_{\mathrm{A}}$ and $P_{\mathrm{M}}$ signals. 
For a given $t$ value, the inverse transformation is
\begin{eqnarray}
M & = & {\mathrm{INT}}[(t+1-m_{\mathrm{t}})/30]+1  \label{month} \\
D & = & t - m_{\mathrm{t}} + 1 - 30(M-1),          \label{day}
\end{eqnarray}
where ${\mathrm{INT}}$ removes the decimal part of $(t+1-m_{\mathrm{t}})/30$.
In other words, if the analysis our data gives any particular $t$ value,
the $D$ and $M$ values of this $t$ can be solved from 
equations (\ref{month}) and (\ref{day}).

The time points $t_{\mathrm{i}}$ for all dates 
with a ``GGG'' or ``SSS'' prognosis combination in CC are
given in Table \ref{table4}. 
These $t_{\mathrm{i}}$ are needed in computing
the binomial distribution probabilities $Q_{\mathrm{B}}$
of equation (\ref{qb}).

\begin{table}[!ht]
\caption{
{\bf The time points $t_{\mathrm{i}}$ of all GGG and SSS dates in CC.}}
\begin{tabular}{|r|r|r|c|}
\hline
$D$  & $M$& $t_{\mathrm{i}}$ & Prog \\
\hline
  1  & 1 &  0.33 & GGG \\
\hline
  2  & 1 &  1.33 & GGG \\
\hline
  5  & 1 &  4.33 & GGG \\
\hline
  7  & 1 &  6.33 & GGG \\
\hline
  9  & 1 &  8.33 & GGG \\
\hline
 10  & 1 &  9.33 & GGG \\
\hline
 11  & 1 & 10.33 & SSS \\
\hline
 12  & 1 & 11.33 & SSS \\
\hline
 16  & 1 & 15.33 & SSS \\
\hline
 17  & 1 & 16.33 & SSS \\
\hline
\end{tabular}
\begin{flushleft} 
The day ($D$) and month ($M$) values in CC 
used in computing the time points
($t_{\mathrm{i}}$) for the days with 
the prognosis (Prog) combinations  ``GGG'' or ``SSS''.
There are $N_{\mathrm{G}}=177$ and $N_{\mathrm{S}}=105$ days
with a ``GGG'' and ``SSS'' combination, respectively.
These data are from Table 1 in Jetsu et al. 2013\cite{Jet13}.
The ten first lines of all 282 lines are shown here for guidance 
regarding the contents of this ASCII file which
can be downloaded on Dryad (http://dx.doi.org/10.5061/dryad.tj4qg).
\end{flushleft}
\label{table4}
\end{table}

\sivu

\section*{Methods}

Let us assume that time is a straight line,
where events are equidistant dots
with a separation of $2 \pi$.
If this line is wound on a $d=1$ diameter wheel,
the dots line up at the same point on the wheel.
Removing some dots produces gaps in the time line, 
but the remaining dots will still line up on the wheel.
However, they will not line up on a $d \neq 1$ diameter wheel.
This is an analogy for the Rayleigh test.
It projects time points on a unit circle with the tested period $P$.
These points line up in the same direction, 
if their time distribution is regular with the tested $P$.

\subsection*{Analysis}

If the Rayleigh method discovers the period $P$
in a series of time points points {\bf t}$=[t_1,t_2, ..., t_{\mathrm{n}}]$,
it is possible to identify those subsamples {\bf t}$^*$ of $n^*$ time points
that strengthen this signal.
In other words, the signal can be separated from noise.
The phases of the $n$ time points $t_{\mathrm{i}}$ are
\begin{eqnarray} 
\phi_{\mathrm{i}}={\mathrm{FRAC}}[(t_{\mathrm{i}}-t_0)/P], 
\label{phase}
\end{eqnarray}
where $t_0$ is an arbitrary zero epoch and ${\mathrm{FRAC}}$
removes the integer part of $(t_{\mathrm{i}}-t_0)/P$.
The unit vectors are 
\begin{eqnarray}
{\bf r}_{\mathrm{i}}=[\cos{\Theta_{\mathrm{i}}},\sin{\Theta_{\mathrm{i}}}], 
\label{unit}
\end{eqnarray}
where $\Theta_{\mathrm{i}}= 360\aste ~ \phi_{\mathrm{i}} $ are the phase angles.
The test statistic of the Rayleigh test is
\begin{eqnarray}
z=|{\bf R}|^2/n, 
\label{rayleigh}
\end{eqnarray}
where vector ${\bf R} =\sum_{\mathrm{i=1}}^{\mathrm{n}} {\bf r}_{\mathrm{i}}$
points to $\Theta_{\mathrm{R}}={\mathrm{atan}}(R_{\mathrm{y}}/R_{\mathrm{x}})$,
$R_{\mathrm{x}} = \sum_{\mathrm{i=1}}^{\mathrm{n}} \cos{\Theta}_{\mathrm{i}}$
and 
$R_{\mathrm{y}}= \sum_{\mathrm{i=1}}^{\mathrm{n}} \sin{\Theta}_{\mathrm{i}}$.
The corresponding phase is $\phi_{\mathrm{R}}=\Theta_{\mathrm{R}}/(360\aste)$.
Coinciding directions $\Theta_\mathrm{i}$ give $|{\bf R}|=n$,
while random $\Theta_\mathrm{i}$ give $|{\bf R}| \approx 0$.
The critical level (i.e. significance) of the Rayleigh test is
\begin{eqnarray}
Q_{\mathrm{z}}={\mathrm{e}}^{-z}. 
\label{qz}
\end{eqnarray}
\noindent
We use the ephemeris zero epoch
\begin{eqnarray}
t_{\mathrm{E}}=t_0 + P \phi_{\mathrm{R}}.
\label{epoch}
\end{eqnarray}
The mutual directions of {\bf r}$_{\mathrm{i}}$ and the length $|{\bf R}|$
are invariant for any constant shift of 
$m_{\mathrm{t}}$, $t_{\mathrm{i}}$, $t_0$ or $t_{\mathrm{E}}$.
Using the above $t_{\mathrm{E}}$ of equation (\ref{epoch}),
vector ${\bf R}$ points to $\Theta=\Theta_R=0\aste$.
All {\bf r}$_i$ with $-90\aste < \Theta_{\mathrm{i}} < 90\aste$
strengthen the $P$ signal, 
while the remaining {\bf r}$_{\mathrm{i}}$ weaken it.
The test statistic can be divided into
$z=R_{\mathrm{x}}^2/n+R_{\mathrm{y}}^2/n$.
We fix $t_0=t_{\mathrm{E}}$ in equation (\ref{phase})
and compute the ``impact'' of any subsample {\bf t}$^*$
on the $P$ signal from
\begin{eqnarray}
z_{\mathrm{x}}= (R_{\mathrm{x}}/|R_{\mathrm{x}}|)(R_{\mathrm{x}}^2/n),
\label{zx}
\end{eqnarray}
where $R_{\mathrm{x}}$ is computed only for the $n=n^*$
time points of {\bf t}$^*$.
These {\bf t}$^*$ may strengthen  ($z_{\mathrm{x}}>0$) 
or
weaken  ($z_{\mathrm{x}}<0$)
the $P$ signal, 
or represent noise ($z_{\mathrm{x}} \approx 0$).

Using the zero epoch $t_0=0$ for 
the $n=6 \times 564$ time points $t_{\mathrm{i}}$ 
of the G prognoses in Table \ref{table3} 
gives the $t_{\mathrm{E}}$ values of Table \ref{table5} for 
the $P_{\mathrm{A}}$ and $P_{\mathrm{M}}$ signals with equation (\ref{epoch}). \\
\begin{table}[!ht]
\caption{
{\bf Values of $t_{\mathrm{E}}$ of the six samples}}
\begin{tabular}{|c|c|c|c|c|c|c|c|} 
\hline
  $P$  &  SSTP=1 & SSTP=3  & SSTP=5   & SSTP=7 & SSTP=9  & SSTP=11 \\
\hline
  2.85 &   0.45  &   0.45  &   0.44  &   0.61  &   0.61  &   0.60  \\
  29.6 &   3.42  &   3.42  &   3.42  &   3.58  &   3.58  &   3.58  \\
\hline
\end{tabular}  
\begin{flushleft} 
\end{flushleft}
\label{table5}
\end{table}
\noindent
These six large samples have
$t_{\mathrm{E}}=0.53\pm0.09$ for $P_{\mathrm{A}}$ and $t_{\mathrm{E}}=3.50\pm0.09$ for $P_{\mathrm{M}}$.
Hence, we use the following two ephemerides
\begin{eqnarray}
t_0=t_{\mathrm{E}}=0.53, &  P = & P_{\mathrm{A}}=2.85 {\mathrm{~~~(Algol)}} \label{aephe}\\
t_0=t_{\mathrm{E}}=3.50, &  P = & P_{\mathrm{M}}=29.6  {\mathrm{~~~(Moon).}}  \label{mephe}
\end{eqnarray}
for computing the phases $\phi_{\mathrm{i}}$ of equation (\ref{phase}).
The lucky ``GGG''
prognoses of every SW are a subsample of the above six large samples
of all ``G'' prognoses.
We give the $z$ and $z_{\mathrm{x}}$ values 
of equations (\ref{rayleigh}) and (\ref{qz})
for any particular SW,
if the analysed $t_{\mathrm{i}}$ of this SW reach $Q_{\mathrm{z}} \le 0.2$ 
with the ephemerides of equations (\ref{aephe}) or (\ref{mephe}).
These periodicities are called weak if $0.05 < Q_{\mathrm{z}} \le 0.2$.

In our Figs. \ref{Horus}--\ref{Man},
we project the $t_{\mathrm{i}}$ of each SW to  
{\bf r}$_{\mathrm{i}}=[\cos{\Theta_{\mathrm{i}}},\sin{\Theta_{\mathrm{i}}}]$ 
on a unit circle,
where time runs in the counter clock--wise direction.
For the $P_{\mathrm{A}}$ signal,
we define four points Aa, Ab, Ac and Ad.
The first one, Aa, is at $\phi=0\equiv0 \aste$ with 
the ephemeris of equation (\ref{aephe}).
The next three points Ab, Ac and Ad are separated by 
$\Delta \phi=0.25 \equiv 90\aste$.
Vectors  {\bf r}$_{\mathrm{i}}$  pointing between Ad $\equiv -90\aste$ and Ab $\equiv +90\aste$
give $z_{\mathrm{x}}>0$ and strengthen $P_{\mathrm{A}}$ signal,
the other ones weaken it.
Because $P_{\mathrm{A}}$ equals $57^{\mathrm{d}}/20$,
the $\phi_{\mathrm{i}}$ of $t_{\mathrm{i}}$ separated 
by multiples of 57 days are equal.  
For clarity, we shift such overlapping $\phi_{\mathrm{i}}$ values by 
$\Delta \phi= 0.005$ away from each other in our Figs. \ref{Horus} -- \ref{Man}. 
However, there are no such shifts in our computations. 
Our unambiguous terminology is: \\ ~ \\
{\it ``Connected to the $P_{\mathrm{A}}$ signal''} $\equiv$  
$t_{\mathrm{i}}$ of an SW strengthen the $P_{\mathrm{A}}$ signal $\equiv$
$z_{\mathrm{x}} \ge 1.0$ and $Q_{\mathrm{z}} \le 0.2$
with the ephemeris of equation (\ref{aephe}). \\
{\it ``Connected to Algol''} $\equiv$ 
$t_{\mathrm{i}}$ of an SW show periodicity with $P_{\mathrm{A}}$, 
but their contribution to the $P_{\mathrm{A}}$ signal 
is insignificant when $0 \le z_{\mathrm{x}} <1.0$
or they weaken this signal when $z_{\mathrm{x}}<0$
$\equiv$
$z_{\mathrm{x}} < 1.0$ and $Q_{\mathrm{z}} \le 0.2$
with the ephemeris of equation (\ref{aephe}). \\ ~ \\

\noindent
We use similar terminology for the {\val Moon} (equation (\ref{mephe})),
and Ma--Md  points similar to Aa--Ad.

Our notations for the lucky and unlucky time points $t_{\mathrm{i}}$ 
of each SW are $g_{\mathrm{i}}$ and $s_{\mathrm{i}}$.
The notations for their unit vectors {\bf r}$_{\mathrm{i}}$ 
of equation (\ref{unit})
are {\bf g}$_{\mathrm{i}}$ and {\bf s}$_{\mathrm{i}}$, respectively.
The critical level $Q_{\mathrm{z}}$ measures the probability for
the concentration of {\it all} $n_{\mathrm{G}}$ and $n_{\mathrm{S}}$
directions of {\bf g}$_{\mathrm{i}}$ and {\bf s}$_{\mathrm{i}}$ of each SW.
These directions are embedded within the directions of
{\it all} {\bf g}$_{\mathrm{i}}$ (Table \ref{table4}: $N_{\mathrm{G}}=177$) 
and {\bf s}$_{\mathrm{i}}$ (Table \ref{table4}: $N_{\mathrm{S}}=105$).
We first choose the direction $\Theta_{\mathrm{R}}$ of ${\bf R}$ for some SW. 
Then we identify the $n_1$ directions of 
{\bf g}$_{\mathrm{i}}$ or {\bf s}$_{\mathrm{i}}$ 
of this SW that 
are among the $n_2$ of all $N_{\mathrm{G}}$ or $N_{\mathrm{S}}$ 
directions closest to $\Theta_{\mathrm{R}}$.
For each SW,
this gives the binomial distribution probability 
\begin{eqnarray}
Q_{\mathrm{B}}=P(n_1,n_2,N)= \sum_{i=n_1}^{n_2} 
{n_2 \choose i} q_{\mathrm{B}}^{i} (1-q_{\mathrm{B}})^{n_2-i}, 
\label{qb}
\end{eqnarray} 
where $N=N_{\mathrm{G}}$ or $N_{\mathrm{S}}$,
and $q_{\mathrm{B}}=n_{\mathrm{G}}/N_{\mathrm{G}}$ or $n_{\mathrm{S}}/N_{\mathrm{S}}$.
This $Q_{\mathrm{B}}$ is the probability for that
the directions of a particular SW occur $n_1$ times, or more, 
among all $n_2$ directions closest to $\Theta_{\mathrm{R}}$.
Many $Q_{\mathrm{z}}$ estimates based on small samples
($n_{\mathrm{G}}$ or $n_{\mathrm{S}}$) are unreliable,
but the $Q_{\mathrm{B}}$ estimates based on 
large samples (Table \ref{table4}: $N_{\mathrm{G}}=177$ or $N_{\mathrm{S}}=105$) are not.

\noindent
All results of our analysis are given in S1 Table, 
where the results mentioned in text are marked with bold letters.
The structure of S1 Table resembles the four panel 
structure of Figs \ref{Horus}--\ref{Man}.
We give four separate tables for each SW.
The results for the lucky and unlucky prognoses with $P_{\mathrm{A}}$ 
are those shown in figure panels ``a'' and ``b''.
The corresponding results for $P_{\mathrm{M}}$ are 
shown in figure panels ``c'' and ``d''.

\section*{Results} 

\subsection*{Algol in lucky prognoses }
\label{Alucky}

Of all 28 SWs, 
only the lucky prognoses of 
{\val Horus}, 
{\val Re}, 
{\val Wedjat},
{\val Followers}, 
{\val Sakhmet}
and
{\val Ennead}
unambiguously strengthen the $P_{\mathrm{A}}$ signal of {\val Algol},
because they have an impact of $z_{\mathrm x} \ge 1.0$ and a significance
of $Q_{\mathrm{z}} \le 0.2$ with the ephemeris of equation (\ref{aephe}).
The lucky prognoses
of {\val Heliopolis} and {\val Enemy} are connected to
{\val Algol} ($Q_{\mathrm{z}} \le 0.2$), 
but they are not connected to the $P_{\mathrm{A}}$ signal ($z_{\mathrm{x}}<1.0$).
In this section, we discuss these eight SWs in the order of their
impact on the $P_{\mathrm{A}}$ signal, i.e. in the order of decreasing
$z_{\mathrm{x}}$ with the ephemeris of equation (\ref{aephe}).

\begin{figure}[h]
\resizebox{12.5cm}{!}{\includegraphics{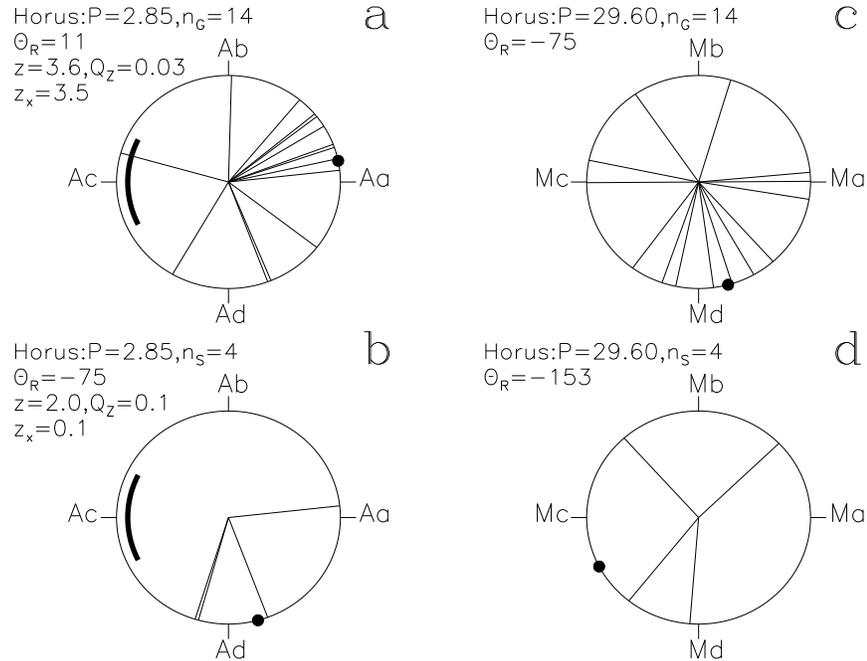}} 
\caption{{\bf Horus.}
Time runs in the counter clock--wise direction on these unit circles.
We give the $z$, $Q_{\mathrm{z}}$
and $z_{\mathrm{x}}$ values only when $Q_{\mathrm{z}} \le 0.2$.
The large black point indicates the $\Theta_{\mathrm{R}}$ direction.
(a) {\bf g}$_{\mathrm{i}}$  with equation (\ref{aephe}).
Point Aa is at $\phi=0\equiv 0\aste$.
The thick line centered on point 
Ac at $\phi=0.5 \equiv 180\aste$ outlines the proposed phase for the 10~hr eclipse of Algol.
(b) {\bf s}$_{\mathrm{i}}$  with equation (\ref{aephe}).
(c) {\bf g}$_{\mathrm{i}}$  with equation (\ref{mephe}).
Point Ma at $\phi=0 \equiv 0\aste$ is close 
to the proposed Full {\val Moon} phase. 
(d) {\bf s}$_{\mathrm{i}}$  with equation (\ref{mephe})
}
\label{Horus}
\end{figure}

\paragraph{Horus} 
This SW has the largest impact $z_{\mathrm{x}}=+3.5$ 
on the $P_{\mathrm{A}}$ signal
and the highest significance of the above eight SWs
($Q_{\mathrm{z}}=0.03$, $n_{\mathrm{G}}=14$).
The unit vectors {\bf g}$_{\mathrm{i}}$ and {\bf s}$_{\mathrm{i}}$ 
of lucky and unlucky prognoses 
with the ephemeris of equation (\ref{aephe}) are shown in Figs \ref{Horus}ab.
Point Aa is at $\phi=0 \equiv 0\aste$.
Points Ab, Ac and Ad are separated by $\Delta \phi=0.25 \equiv 90\aste$.
Only the {\bf g}$_{\mathrm{i}}$ pointing 
between Ad $\equiv -90\aste$ and Ab $\equiv +90\aste$ 
strengthen the $P_{\mathrm{A}}$ signal.
Twelve out of all fourteen {\bf g}$_{\mathrm{i}}$ 
are within this interval (Fig \ref{Horus}a).
The four $\Theta_{\mathrm{i}}$ closest to $\Theta_{\mathrm{R}}=11\aste$ reach 
a high significance of
$Q_{\mathrm{B}}=0.006$ ($n_1=4$, $n_2=10$, $N_{\mathrm{G}}=177$).
The {\bf g}$_{\mathrm{i}}$ pointing closest to Aa and
giving the strongest impact on the $P_{\mathrm{A}}$ signal
has the CC text\cite{Bak66}  \\ 
$g_{\mathrm{i}}(14,2)\equiv +6\aste$:
{\it
``It is the day of receiving the white crown 
by the Majesty of {\val Horus}; his Ennead is in great festivity.''
} \\ 
The texts\cite{Bak66,Lei94} for the next best 
{\bf g}$_{\mathrm{i}}$ closest to Aa are \\ 
$g_{\mathrm{i}}(19,12)\equiv +13\aste:$
{\it
``Horus has returned complete, nothing is missing in it.''
} \\ 
$g_{\mathrm{i}}(27,1)\equiv +19\aste:$
{\it
``Peace on the part of Horus with Seth.''
} \\ 
$g_{\mathrm{i}}(24,3)\equiv +19\aste:$
{\it
``He has given his throne to his son, Horus, in front of Re.''
} \\ 
$g_{\mathrm{i}}(1,7)\equiv +32\aste:$
{\it
``Feast of entering into heaven and the two banks. Horus is jubilating.''
} \\ 
$g_{\mathrm{i}}(15,11)\equiv +38\aste:$
{\it
``Horus hears your words in the presence of every god and goddess on this day.''
} \\ 
$g_{\mathrm{i}}(27,3)\equiv +38\aste:$
{\it
``Judging Horus and Seth; stopping the fighting.''
} \\ 
$g_{\mathrm{i}}(18,1)\equiv -38\aste:$
{\it
``It is the day of magnifying the majesty of Horus more than his brother, ...''
} \\ 
$g_{\mathrm{i}}(1,9)\equiv +51\aste:$
{\it
``Feast of Horus son of Isis and ... his followers ... day''
} \\ 
$g_{\mathrm{i}}(23,7)\equiv -69\aste:$
{\it
``Feast of Horus ... on this day of his years in his very beautiful images.''
} \\ 
$g_{\mathrm{i}}(29,3)\equiv -69\aste:$
{\it
``White crown to Horus, and the red one to Seth.''
} \\ 
$g_{\mathrm{i}}(7,9)\equiv +88\aste:$
{\it
``The crew follow Horus in the foreign land, 
examining its list ... therein when he smote 
him who rebelled against his master.''
} \\ 
$g_{\mathrm{i}}(1,10)\equiv -120\aste:$
{\it
``Horus ... Osiris ... Chentechtai ... land ''
} \\ 
$g_{\mathrm{i}}(28,3)\equiv +164\aste:$
{\it
``The gods are in jubilation and in joy when the will is written (lit. made) for Horus, ...''
} \\ 
These passages of lucky prognoses 
are suggestive of {\val Algol} at its brightest. 
The {\it ``white crown''}, {\val Horus} having {\it ``returned complete''} and 
{\it ``entering into heaven''} (i.e. into the sky)
are not easy to explain as symbols for the eclipse.
Among the $g_{\mathrm{i}}$ of all 28 SWs,
the $g_{\mathrm{i}}$ of {\val Horus} are the ``best hit'' 
on Aa ($z_{\mathrm{x}}=+3.5$).
If these $g_{\mathrm{i}}$ represent {\val Algol} at its brightest,
then Aa is in the middle of this brightest phase and
the thick line centered at Ac in Fig \ref{Horus}a
outlines {\val Algol}'s eclipse.
In this case, the $g_{\mathrm{i}}(7,9)\!\equiv\! +88\aste$ 
text may refer to an imminent eclipse and
{\it ``the will is written''} in $g_{\mathrm{i}}(28,3)\!\equiv\! +164\aste$ 
to the moment when the beginning of the 
eclipse is just becoming observable with naked eye.
These passages could certainly describe naked eye 
observations of the regular changes of {\val Algol}.

Three {\bf s}$_{\mathrm{i}}$ of {\val Horus} in Fig \ref{Horus}b
concentrate close to Ad and reach 
$Q_{\mathrm{B}}=0.07$ ($n_1=3$, $n_2=25$, $N_{\mathrm{S}}=105$).
The fourth vector {\bf s}$_{\mathrm{i}}$ points close to Aa. 
Their CC texts\cite{Bak66} are \\
$s_{\mathrm{i}}(26,1)\equiv -107\aste:$
{\it
``... It is the day of Horus fighting with Seth. ...''
} \\ 
$s_{\mathrm{i}}(11,11)\equiv -107\aste:$
{\it
``Introducing the great ones by Re to the booth
to see what he
had observed through the eye of Horus the elder.
They were with heads bent down when they saw the eye of Horus being angry
in front of Re.''
} \\ 
$s_{\mathrm{i}}(20,9)\equiv -69\aste:$
{\it
``Mat judges in front of these gods who became angry in the island of 
the sanctuary of Letopolis. The Majesty of Horus revised it.''
} \\ 
$s_{\mathrm{i}}(5,8)\equiv 6\aste:$
{\it
``The Majesty of Horus is well when the red one sees his form.
As for anybody who approaches it, anger will
start on it.''
} \\ 
If the $g_{\mathrm{i}}$ that described feasts were connected 
to the brightest phase of {\val Algol}, these $s_{\mathrm{i}}$ describing anger 
would have occurred  after {\val Algol}'s eclipse.
{\it ``Horus is well''} for the last $s_{\mathrm{i}}(5,8)$ 
would seem natural for a lucky prognosis of {\val Horus} (as it should be close to Aa) 
but it is deemed unlucky for some other reasons.
This type of ``conflict of interest'' prognoses may explain,
why there are significant concentrations of directions 
accompanied by a few irregular directions
(e.g. Fig \ref{Earth}c).

The $g_{\mathrm{i}}$ and $s_{\mathrm{i}}$ of {\val Horus}
have $Q_{\mathrm{z}}>0.2$ with the ephemeris of equation (\ref{mephe}),
and are therefore not connected to the {\val Moon}, 
except for some $g_{\mathrm{i}}$ 
texts mentioning both {\val Horus} and {\val Seth}.
We argue that, as Leitz\cite{Lei94} also did, 
Mc $\equiv 180\aste$ in Fig \ref{Horus}c coincides with the New {\val Moon}
(see paragraph {\val \nameref{Sethpara}}).
All the aforementioned lucky prognoses 
mentioning both {\val Horus} and {\val Seth}
are close to Md $\equiv -90\aste$ in Fig \ref{Horus}c,
i.e.
$g_{\mathrm{i}}(27,1)\equiv -82\aste$,
$g_{\mathrm{i}}(27,3)\equiv -73\aste$ 
and
$g_{\mathrm{i}}(29,3)\equiv -48\aste$ 
with the ephemeris of equation (\ref{mephe}).
The texts of these three days may describe the
``luminosity competitions'' between {\val Horus} and {\val Seth}
which come to an end when more
than half of the lunar disk becomes illuminated immediately after Md.
The legend of the Contendings of 
{\val Horus} and {\val Seth}\cite{Lic76} (hereafter LE1)
has inspired these descriptions.
The text
{\it ``White crown to Horus, and the red one to Seth''}
in $g_{\mathrm{i}}(29,3)$
would describe the brightening of {\val Horus} with {\val Algol} 
(Fig. \ref{Horus}a: $\Theta\!=\!-69\aste$)
and the brightening of {\val Seth} 
(Fig. \ref{Horus}c: $\Theta\!=\!-48\aste$)
with the approaching Full {\val Moon} at Ma.
The most simple explanation for the context of these texts
is that the lucky prognoses of {\val Horus} are connected
to {\val Algol} at its brightest.

\begin{figure}[h]
\resizebox{12.5cm}{!}{\includegraphics{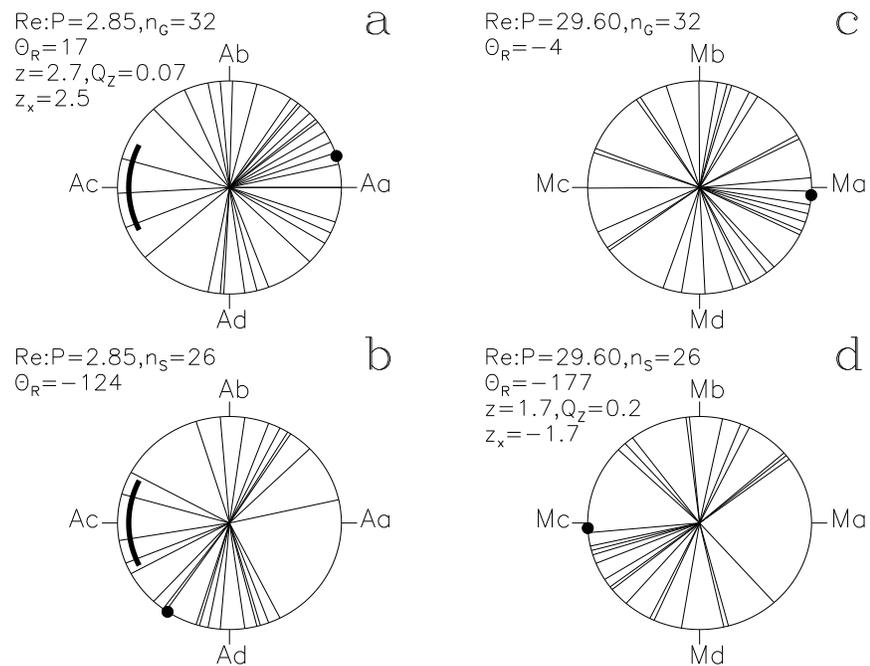}} 
\caption{{\bf Re.} otherwise as in Fig \ref{Horus}}
\label{Re}
\end{figure}

\paragraph{Re} 
The lucky prognoses reach $Q_{\mathrm{z}}=0.07$ ($n_{\mathrm{G}}=32$)
with the ephemeris of equation (\ref{aephe}) and
give the second largest impact $z_{\mathrm{x}}=+2.5$ on 
the $P_{\mathrm{A}}$ signal
(Fig \ref{Re}a).
Absence of small $Q_{\mathrm{B}}$ values,
i.e. {\bf g}$_{\mathrm{i}}$ concentrations,
may indicate that {\val Re} (the {\val Sun})  was casually 
following the undertakings of {\val Horus}.
The {\bf s}$_{\mathrm{i}}$ of {\val Re} 
reach $Q_{\mathrm{z}}=0.2$ ($n_{\mathrm{S}}=26$) with the ephemeris of
equation (\ref{mephe}), and explicitly
avoid Ma, the proposed Full {\val Moon} phase
(Fig \ref{Re}d).

\sivu

\begin{figure}[h]
\resizebox{12.5cm}{!}{\includegraphics{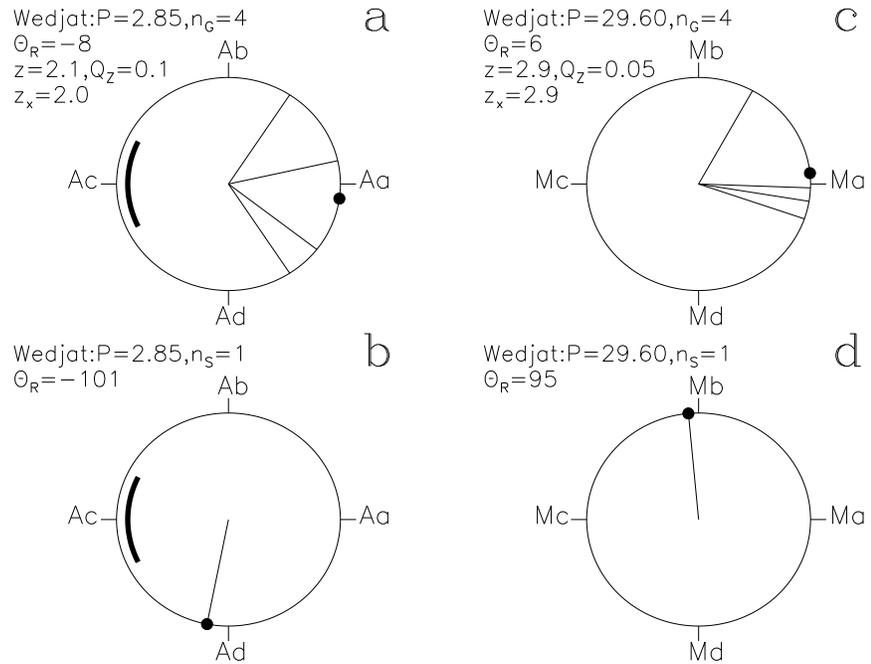}} 
\caption{{\bf Wedjat.} otherwise as in Fig \ref{Horus}}
\label{Wedjat}
\end{figure}

\paragraph{Wedjat} The lucky prognoses show weak periodicity 
($Q_{\mathrm{z}}=0.1$, $n_{\mathrm{G}}=4$)
with the
ephemeris of equation 
(\ref{aephe}). They give the third largest impact 
$z_{\mathrm{x}}=+2.0$ on the $P_{\mathrm{A}}$ signal
(Fig \ref{Wedjat}a).
However, their impact on the $P_{\mathrm{M}}$ signal is even larger,
$z_{\mathrm{x}}=+2.9$ (Fig \ref{Wedjat}c).
{\val Wedjat} may represent {\val Algol} observed
at its brightest close to the Full {\val Moon}.
The {\bf g}$_{\mathrm{i}}$ and {\bf s}$_{\mathrm{i}}$ distributions 
of {\val Horus} and {\val Wedjat} are similar 
(Figs \ref{Horus}ab and \ref{Wedjat}ab)
with the ephemeris of equation (\ref{aephe}).
{\val Wedjat} is the Eye of Horus in Ancient Egyptian mythology.

\sivu

\begin{figure}[h]
\resizebox{12.5cm}{!}{\includegraphics{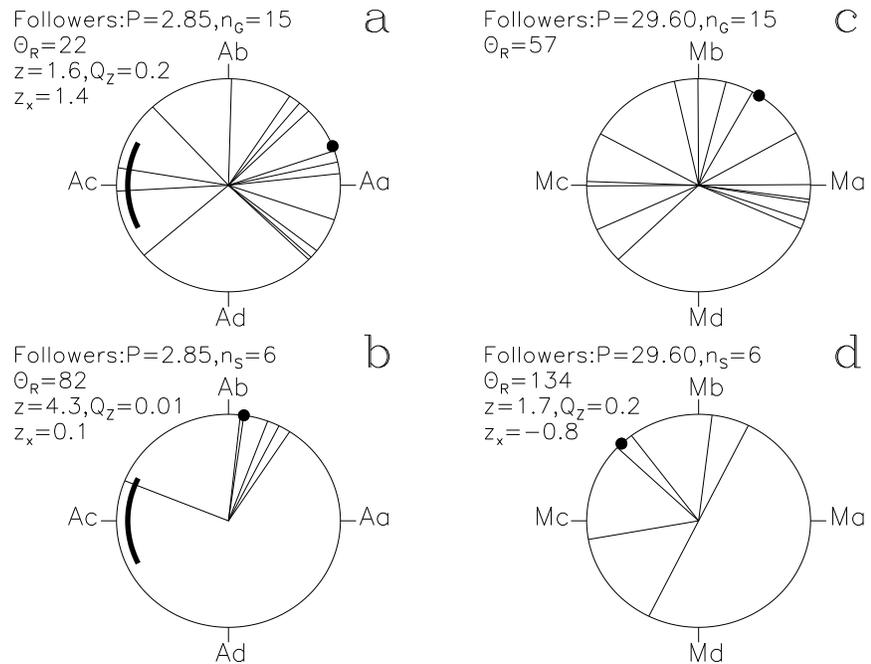}} 
\caption{{\bf Followers.} otherwise as in Fig \ref{Horus}}
\label{Followers}
\end{figure}

\paragraph{Followers} 
The lucky prognoses have an impact of $z_{\mathrm{x}}=+1.4$
on the $P_{\mathrm{A}}$ signal (Fig \ref{Followers}a).
This periodicity is weak ($Q_{\mathrm{z}}=0.2$, $n_{\mathrm{G}}=15$).
Six $s_{\mathrm{i}}$ reach $Q_{\mathrm{z}}=0.01$ (Fig. \ref{Followers}b).
The five {\bf s}$_{\mathrm{i}}$ closest to $\Theta_{\mathrm{R}}$ reach a high
significance of
$Q_{\mathrm{B}}=0.003$ ($n_1=5$, $n_2=18$, $N_{\mathrm{S}}=105$)
and may refer to an approaching eclipse of {\val Algol}.
These $s_{\mathrm{i}}$ also show a weak connection to the {\val Moon}
(Fig. \ref{Followers}d).
It is tempting to suggest that {\val Followers} would be 
{\val Pleiades} following very close behind {\val Algol} in the revolving sky,
e.g. in $g_{\mathrm{i}}(7,9) \equiv 88\aste$ 
{\it ``The crew follow Horus in the foreign land''}
(Figs. \ref{Horus}a and \ref{Followers}a).

\sivu

\begin{figure}[h]
\resizebox{12.5cm}{!}{\includegraphics{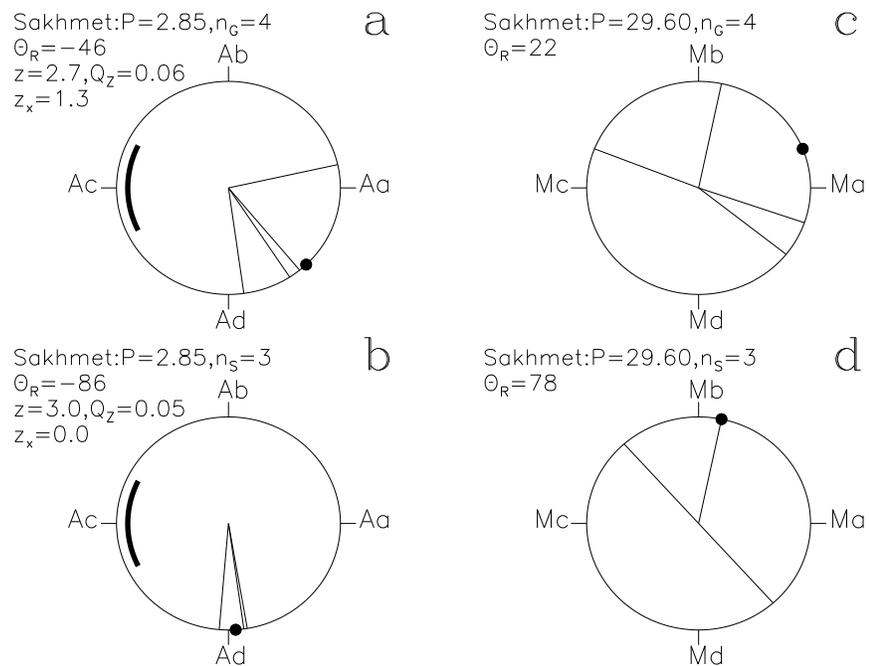}} 
\caption{{\bf Sakhmet.} otherwise as in Fig \ref{Horus}}
\label{Sakhmet}
\end{figure}

\paragraph{Sakhmet}
The $g_{\mathrm{i}}$ and $s_{\mathrm{i}}$ 
reach $Q_{\mathrm{z}}=0.06$ ($n_{\mathrm{G}}=4$)
and 0.05 ($n_{\mathrm{S}}=3$) with the ephemeris of equation (\ref{aephe}).
The impact of  $g_{\mathrm{i}}$ on the $P_{\mathrm{A}}$ signal 
is $z_{\mathrm{x}}=+1.3$
(Fig. \ref{Sakhmet}a).
The three $s_{\mathrm{i}}$ at Ad, 
after the proposed eclipse at Ac, 
are strongly connected to {\val Algol},
because they reach the most extreme significance in this study,
$Q_{\mathrm{B}}=0.0004$ ($n_1=3$, $n_2=6$, $N_{\mathrm{S}}=105$).
The texts\cite{Bak66} are \\ 
$s_{\mathrm{i}}(27,8)\equiv -95\aste:$
{\it ``Re sets because the Majesty of the goddess Sakhmet is angry in the land of Temhu.''} \\
$s_{\mathrm{i}}(13,6)\equiv -82\aste:$
{\it
``It is the day of the proceeding of Sakhmet to Letopolis. 
Her great executioners passed by the offerings of Letopolis on this day.''} \\
$s_{\mathrm{i}}(7,10)\equiv -82\aste:$
{\it
``It is the day of the executioners of Sakhmet.''
} \\ 
\noindent
These three unlucky prognoses (Fig. \ref{Sakhmet}b)
are immediately followed by lucky ones (Fig. \ref{Sakhmet}a).
The {\bf g}$_{\mathrm{i}}$ and {\bf s}$_{\mathrm{i}}$
distributions of {\val Sakhmet} (Fig \ref{Sakhmet}ab) 
resemble those of {\val Horus} (Fig \ref{Horus}ab)
with the ephemeris of equation (\ref{aephe}).
The Eye of Horus ({\val Wedjat}) was transformed into 
the vengeful goddess {\val Sakhmet}
in the legend\cite{Lic76} of the Destruction of Mankind
(hereafter LE2).
The {\bf s}$_{\mathrm{i}}$ vectors of {\val Horus}, 
{\val Wedjat} and {\val Sakhmet} point close to Ad which is after 
{\val Algol}'s proposed eclipse at Ac 
(Figs \ref{Horus}b, \ref{Wedjat}b and \ref{Sakhmet}b), and
may refer to the abrupt pacification of 
enraged {\val Sakhmet} in LE2.

\sivu

\begin{figure}[h]
\resizebox{12.5cm}{!}{\includegraphics{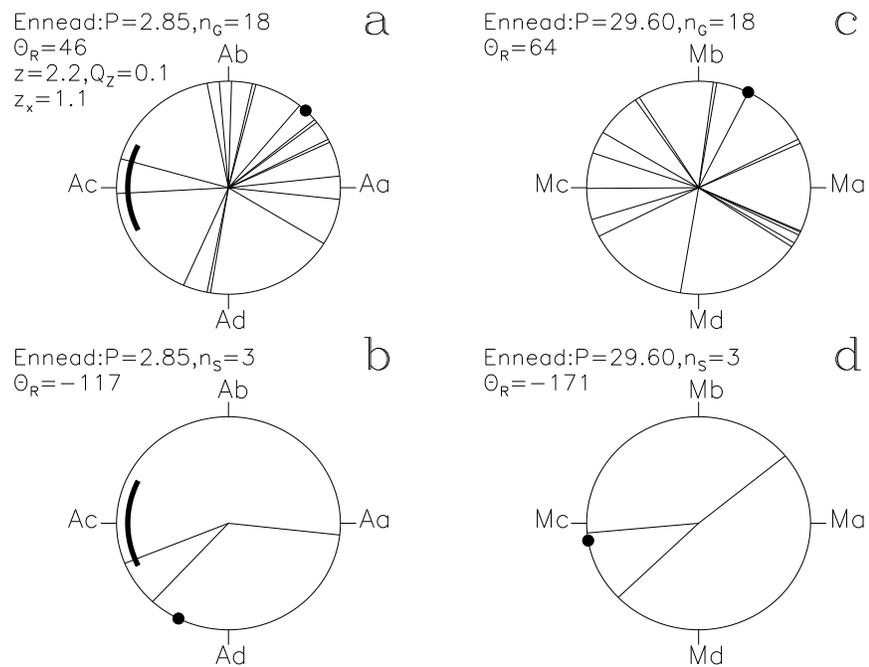}} 
\caption{{\bf Ennead.} otherwise as in Fig \ref{Horus}}
\label{Ennead}
\end{figure}

\paragraph{Ennead}
The lucky prognoses show weak periodicity (Fig \ref{Ennead}a: $Q_{\mathrm{z}}=0.1$,$n_{\mathrm{G}}=18$)
and an impact of $z_{\mathrm{x}}=+1.1$ on the $P_{\mathrm{A}}$ signal
with the ephemeris of equation (\ref{aephe}),
as well as some concentration
($Q_{\mathrm{B}}=0.02$, $n_1=12$, $n_2=63$, $N_{\mathrm{G}}=177$).
Ennead was a group of nine deities in Ancient Egyptian mythology.
We discussed earlier, why {\val Followers} may have represented {\val Pleiades}.
{\val Ennead} may have been another name for {\val Pleiades},
having the modern name ``Seven sisters''.
However, the number of {\val Pleiades} members visible with naked 
eye depends on the observing conditions and the observer,
the maximum number of such members 
being fourteen \cite{Win78,Ada11}.
The unlucky prognoses of {\val Followers} could be connected to 
{\val Pleiades} following the disappearing {\val Algol} 
before eclipse (Fig. \ref{Followers}b),
while the unlucky prognoses of {\val Ennead} could be connected to
{\val Algol} reappearing in front {\val Pleiades} after eclipse (Fig. \ref{Ennead}b).
Furthermore, the lucky prognosis distributions of {\val Followers} and {\val Ennead} 
are very similar (Figs. \ref{Followers}a and \ref{Ennead}a).

\paragraph{Heliopolis} The lucky prognoses show weak periodicity with
${P_{\mathrm{A}}}$, but their impact on this signal is insignificant,
$z_{\mathrm{x}}=+0.2$, with the ephemeris of equation (\ref{aephe}).

\paragraph{Enemy} These lucky prognoses weaken the ${ P_{\mathrm{A}}}$ signal,
because their impact is $z_{\mathrm{x}}=-1.0$
with the ephemeris of equation (\ref{aephe}).

\sivu

\subsection*{The Moon in lucky prognoses}
\label{Mlucky}

We discuss the remaining other 20 SWs in this section and in sections
\begin{itemize}
\item[] \nameref{Aunlucky}
\item[] \nameref{Munlucky}
\item[] \nameref{Noconnection}
\end{itemize}
These SWs are discussed only briefly,
because they are not connected to the $P_ {\mathrm{A}}$ signal.

The lucky prognoses of 
{\val Earth},
{\val Heaven},
{\val Busiris},
{\val Rebel},
{\val Thoth} and 
{\val Onnophris}
are connected to
the ${P_{\mathrm{M}}}$ signal,
because they have $z_{\mathrm{x}} \ge 1.0$ and $Q_{\mathrm{z}} \le 0.2$
with the ephemeris of equation (\ref{mephe}).
The lucky prognoses of
{\val Nut} are weakly connected to the {\val Moon}.

\begin{figure}[h]
\resizebox{12.5cm}{!}{\includegraphics{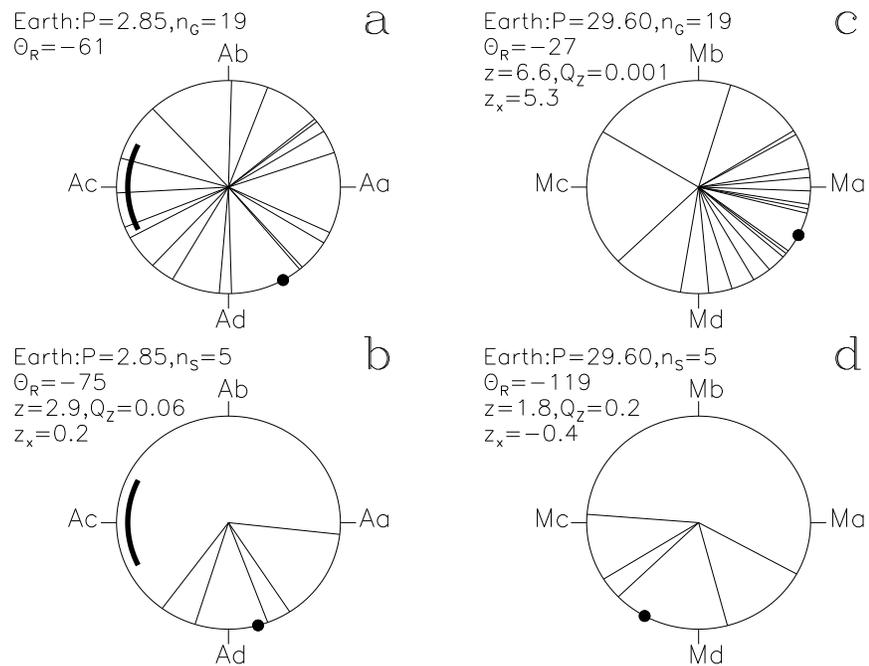}} 
\caption{{\bf Earth.} otherwise as in Fig \ref{Horus}}
\label{Earth}
\end{figure}

\paragraph{Earth} These lucky prognoses reach the highest impact parameter 
value of this study, $z_{\mathrm{x}}=+5.3$, on the $P_{\mathrm{M}}$ signal.
This periodicity also reaches the highest Rayleigh test significance of all,
$Q_{\mathrm{z}}=0.001$ ($n_{\mathrm{G}}=19$).
The good moments on {\val Earth} occurred before and during Ma, the proposed Full {\val Moon}
phase (Fig \ref{Earth}c).
The unlucky prognoses also show a weak connection to {\val Algol}
(Fig. \ref{Earth}b: $Q_{\mathrm{z}}=0.06$, $n_{\mathrm{S}}=5$) and an even weaker
connection to the {\val Moon}
(Fig. \ref{Earth}d: $Q_{\mathrm{z}}=0.2$, $n_{\mathrm{S}}=5$).

\sivu

\begin{figure}[h]
\resizebox{12.5cm}{!}{\includegraphics{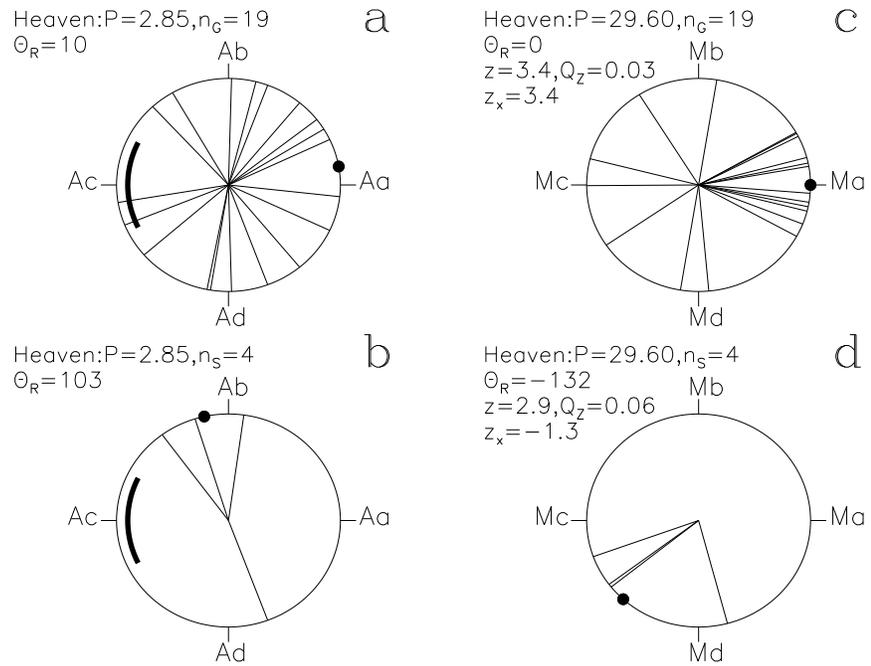}} 
\caption{{\bf Heaven.} otherwise as in Fig \ref{Horus}}
\label{Heaven}
\end{figure}

\paragraph{Heaven} The second largest impact $z_{\mathrm{x}}=+3.4$ 
on the $P_{\mathrm{M}}$ signal comes from these lucky prognoses.
Again, the good moments coincide with Ma, the proposed Full {\val Moon} phase 
(Fig \ref{Heaven}c).
This is significant periodicity 
($Q_{\mathrm{z}}=0.03$, $n_{\mathrm{G}}=19$)
combined with a very significant concentration 
($Q_{\mathrm{B}}=0.002$, $n_1=12$, $n_2=45$, $N_{\mathrm{G}}=177$).
The unlucky prognoses also show a weak connection 
to the {\val Moon} (Fig. \ref{Heaven}d: $Q_{\mathrm{z}}=0.06$, $n_{\mathrm{S}}=4$).

\sivu

\begin{figure}[h]
\resizebox{12.5cm}{!}{\includegraphics{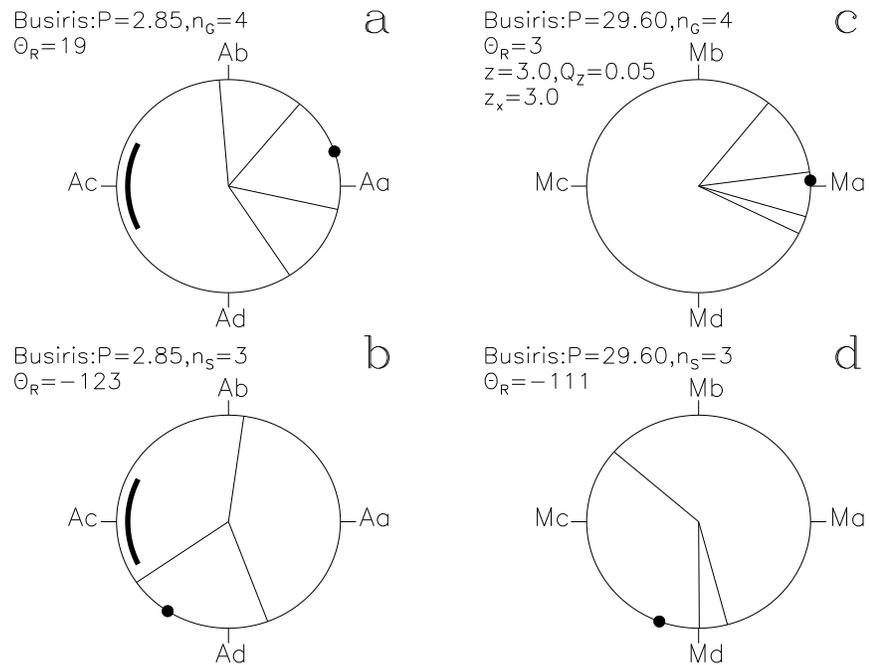}} 
\caption{{\bf Busiris.} otherwise as in Fig \ref{Horus}}
\label{Busiris}
\end{figure}

\paragraph{Busiris} The third largest impact
on the $P_{\mathrm{M}}$ signal,
$z_{\mathrm{x}}=+3.0$, 
comes from the lucky prognoses of {\val Busiris}.
This periodicity reaches $Q_{\mathrm{z}}=0.05$ ($n_{\mathrm{G}}=4$)
with the ephemeris of equation (\ref{mephe}).
And again, the lucky prognoses are close to Ma, 
the proposed Full {\val Moon} phase (Fig \ref{Busiris}c)

\paragraph{Rebel} The lucky prognoses
show weak periodicity ($Q_{\mathrm{z}}=0.2$, $n_{\mathrm{G}}=3$)
with the ephemeris of equation (\ref{mephe}) and have
an impact of $z_{\mathrm{x}} = 1.6$ on the $P_{\mathrm{M}}$ signal.

\paragraph{Thoth and Onnophris} 
The lucky prognoses of these SW have a weaker
impact on the $P_{\mathrm{M}}$ signal, i.e.
$1.0 \le z_{\mathrm{x}} \le 1.3$ with the ephemeris equation (\ref{mephe}).

\paragraph{Nut} The lucky prognoses show a weak 
connection to the {\val Moon}. They have no impact on $P_{\mathrm{M}}$, 
because $z_{\mathrm{x}}=-0.1$ with the ephemeris of equation (\ref{mephe}).

\sivu

\subsection*{Algol in unlucky prognoses}
\label{Aunlucky}

The $P_{\mathrm{A}}$ and $P_{\mathrm{M}}$ signals were detected
from the lucky prognoses  $g_{\mathrm{i}}$ \cite{Por08,Jet13}. 
It is therefore self--evident that the unlucky prognoses $s_{\mathrm{i}}$
had no impact on these two signals.  
However, this does not rule out the possibility that
the $s_{\mathrm{i}}$ of some SW may be connected to {\val Algol} or the {\val Moon}.
Most of these {\bf s}$_{\mathrm{i}}$ vectors point away 
from Aa or Ma, i.e. $z_{\mathrm{x}}<0$ with 
the ephemerides of equations 
(\ref{aephe}) or (\ref{mephe}).
{\val Man} and {\val Flame} are 
the only exceptions to this general rule ($z_{\mathrm{x}} \ge 0$).

\begin{figure}[h]
\resizebox{12.5cm}{!}{\includegraphics{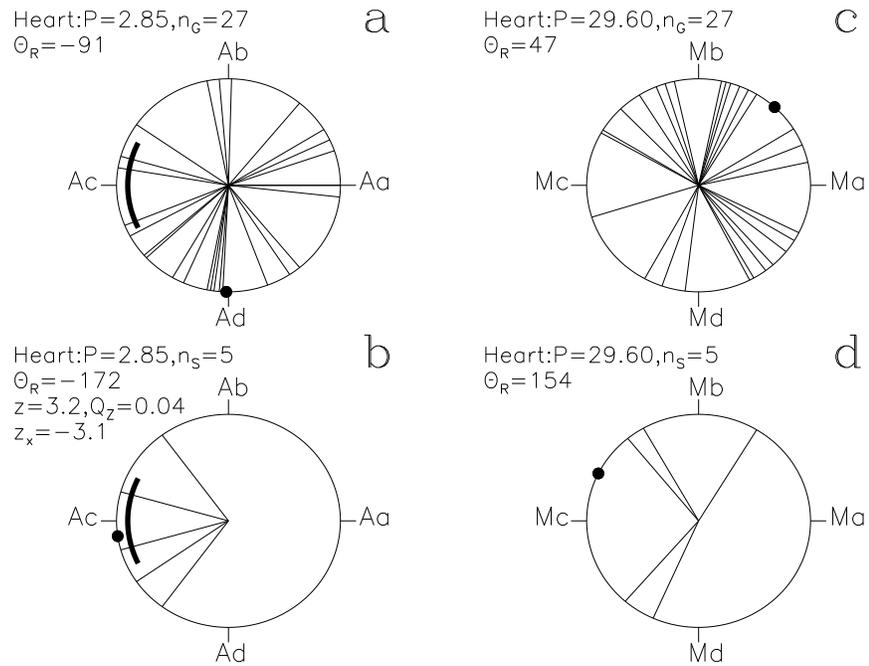}} 
\caption{{\bf Heart.} otherwise as in Fig \ref{Horus}}
\label{Heart}
\end{figure}

\paragraph{Heart} The unlucky prognoses have 
$z_{\mathrm{x}}=-3.1$ with the ephemeris of equation (\ref{aephe}).
They point towards Ac, the proposed eclipse phase of {\val Algol} 
(Fig \ref{Heart}b).
This periodicity reaches a significance of
$Q_{\mathrm{z}}=0.04$ ($n_{\mathrm{S}}=5$)
and $Q_{\mathrm{B}}=0.04$ ($n_1=5$, $n_2=39$, $N_{\mathrm{S}}=105$).

\paragraph{Nun} The three unlucky prognoses of this SW
reach $Q_{\mathrm{z}}=0.06$ and a high significance of 
$Q_{\mathrm{B}}=0.003$ ($n_1=3$, $n_2=11$, $N_{\mathrm{S}}=105$)
with the ephemeris of equation (\ref{aephe}).
They also show a weaker connection to the {\val Moon}.

\sivu

\subsection*{The Moon in unlucky prognoses}
\label{Munlucky}

We will first discuss the unlucky prognoses of
SWs having negative $z_{\mathrm{x}}$ values
with the ephemeris of equation (\ref{mephe}), and then the two exceptions 
of {\val Man} and {\val Flame}.

\begin{figure}[h]
\resizebox{12.5cm}{!}{\includegraphics{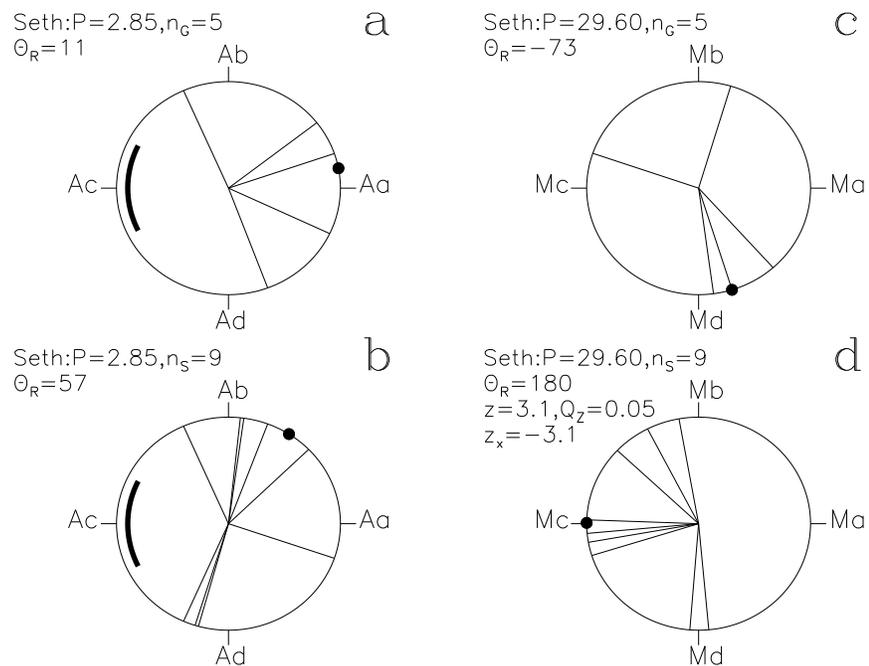}} 
\caption{{\bf Seth.} otherwise as in Fig \ref{Horus}}
\label{Seth}
\end{figure}

\paragraph{Seth} 
\label{Sethpara}
``See you on the dark side of the Moon'' sums up
the unlucky prognoses of {\val Seth} (Fig \ref{Seth}d).
The significance is $Q_{\mathrm{z}}=0.05$ ($n_{\mathrm{S}}=9$)
with the ephemeris of equation (\ref{mephe}).
Leitz\cite{Lei94} has argued that the following texts\cite{Bak66} 
at two consecutive days \\
$s_{\mathrm{i}}(16,7)\equiv 173\aste$: 
{\it 
``Do not look, darkness being on this day 
(or, do not see darkness on this day).''} \\
$s_{\mathrm{i}}(17,7) \equiv 185\aste$: 
{\it ``Do not pronounce the name of Seth on this day.''} \\
\noindent
take place during the New {\val Moon}. 
The {\bf s}$_{\mathrm{i}}$ vectors of these 
two particular texts point at the opposite sides of Mc $\equiv 180\aste$, 
which supports both our ``prediction'' formula of equation (\ref{mephe}) 
and Leitz' attribution\cite{Lei94} of the texts to the New {\val Moon}.
We conclude that {\val Seth} is connected to the {\val Moon}  and 
strongly suggest that Mc computed with 
equation (\ref{mephe}) is close to the New {\val Moon}.
Hence, the Full {\val Moon} is close to Ma.

\sivu

\begin{figure}[h]
\resizebox{12.5cm}{!}{\includegraphics{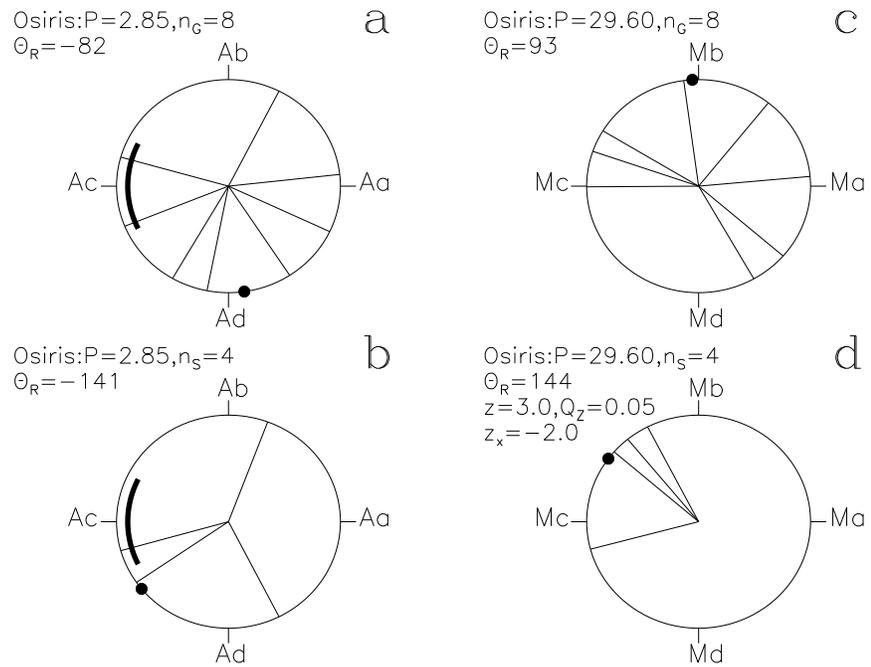}} 
\caption{{\bf Osiris.} otherwise as in Fig \ref{Horus}}
\label{Osiris}
\end{figure}

\paragraph{Osiris} The four unlucky prognoses of this SW also point
to the dark side of the {\val Moon}, 
assuming that Mc is close to the New {\val Moon} 
(Fig \ref{Osiris}d). 
The significance estimates are $Q_{\mathrm{z}}=0.05$ ($n_{\mathrm{S}}=4$)
and $Q_{\mathrm{B}}=0.02$ ($n_1=3$, $n_2=15$, $N_{\mathrm{S}}=105$)
with the ephemeris of equation (\ref{mephe}).

\paragraph{Abydos and Lion} These unlucky prognoses
show a weak connection to the {\val Moon}.

\sivu

\begin{figure}[h]
\resizebox{12.5cm}{!}{\includegraphics{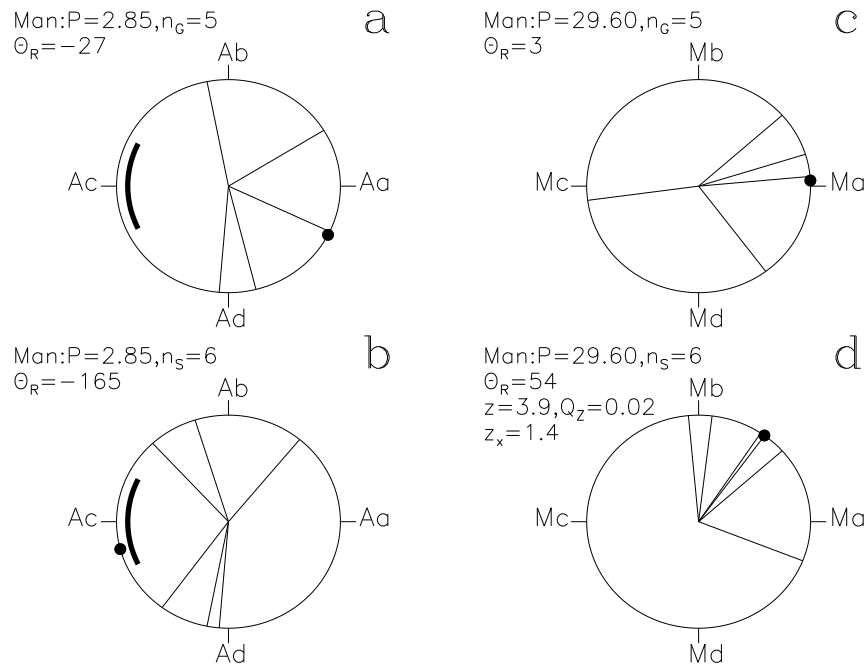}} 
\caption{{\bf Man.} otherwise as in Fig \ref{Horus}}
\label{Man}
\end{figure}

\paragraph{Man} The significance estimates for 
the unlucky prognoses are
$Q_{\mathrm{z}}=0.02$  ($n_{\mathrm{S}}=6$)
and $Q_{\mathrm{B}}=0.009$ ($n_1=5$, $n_2=23$, $N_{\mathrm{S}}=105$)
with the ephemeris equation (\ref{mephe}).
These unlucky moments of {\val Man} 
concentrate on a few days after Ma, the proposed Full {\val Moon} 
phase (Fig \ref{Man}d).

\paragraph{Flame} The significance estimates for these unlucky prognoses are
$Q_{\mathrm{z}}=0.03$ ($n_{\mathrm{S}}=4$) and  
$Q_{\mathrm{B}}=0.003$ ($n_1=4$, $n_2=17$, $N_{\mathrm{S}}=105$)
with the ephemeris of equation (\ref{mephe}).

\subsection*{No Algol or the Moon in lucky or unlucky prognoses}
\label{Noconnection}

\paragraph{Eye, Fire, Majesty, Shu and Sobek}
These SWs are not connected to {\val Algol} or the {\val Moon},
because their $g_{\mathrm{i}}$ and  $s_{\mathrm{i}}$
have $Q_{\mathrm{z}}>0.2$ with the ephemerides of
equations (\ref{aephe}) and (\ref{mephe}). 

\subsection*{Some general remarks}
This concludes our analysis of 28 SWs.
Numerous other\cite{Har02} SWs in CC
need to be analysed in the future.
Combining the inverse relations 
of equations (\ref{month}) and (\ref{day})
to the ephemerides of equations (\ref{aephe}) and (\ref{mephe}) 
will have countless applications.
For example,
the first eclipse of {\val Algol} would have occurred on
$t(2.6,1)=1.96$ at $D=2.1$ in $M=1$
or the last New {\val Moon} on
$t(14.6,12)=343.9$ at $D=14.6$ in $M=12$.
Any question about CC
can now be studied within this precise framework, e.g.
was some meaning given to the nights
when an eclipse of {\val Algol} (equation (\ref{aephe}): $\phi=0.5$)
coincided with the New {\val Moon}   (equation (\ref{mephe}): $\phi=0.5$)?

\sivu

\section*{Discussion}

Previously,
we\cite{Jet13} applied four tests to the astrophysical hypothesis \\
$H_1:$ {\it ``Period $P_{\mathrm{A}}=2.^{\mathrm{d}}850$ in CC was $P_{\mathrm{orb}}$ of Algol.''} \\
\noindent
This is a summary of those tests: \\
{\sc test i}: The mass transfer in this binary system
should have increased the period in the past three millennia. 
The period value in CC is the first evidence for such an increase
since Goodricke\cite{Goo83} discovered this periodicity over two centuries ago. \\
{\sc test ii}: The period change of 0.017 days from 2.850 to 2.867 days
gives a reasonable estimate for the rate of this mass transfer. \\
{\sc test iii}: If eclipses were observed in Ancient Egypt,
the orbital plane of the Algol A--B system must
be nearly perpendicular to that of the Algol AB--C system\cite{Zav10,Bar12}. \\
{\sc test iv}: {\val Algol} and the {\val Moon} are
the most probable objects, where naked
eye observers could have discovered periodicity 
that we could then rediscover in CC. \\

{\sc tests i} and {\sc iv} supported $H_1$, 
while {\sc tests ii} and {\sc iii} indicated that it could be true.

\begin{figure}[h]
\resizebox{12.5cm}{!}{\includegraphics{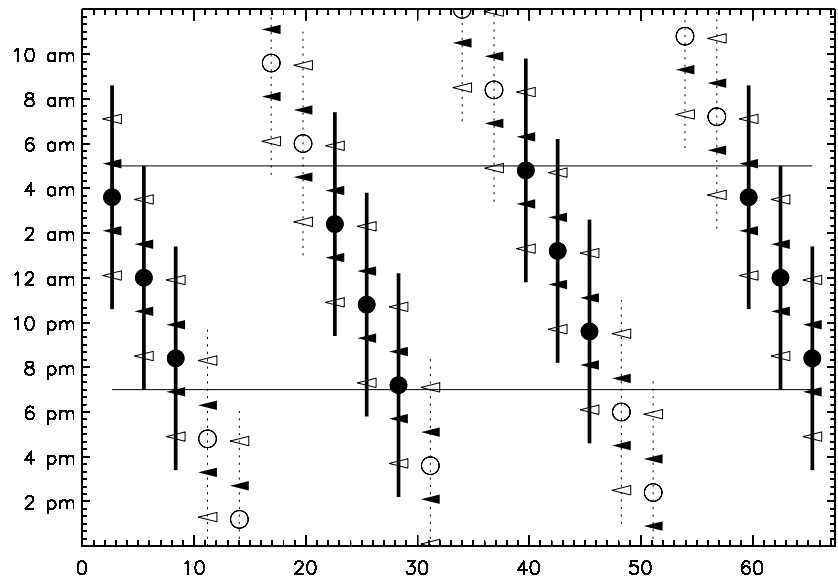}} 
\caption{{\bf Eclipses of Algol with $P_{\mathrm{A}}=2.85$ days.} 
The horizontal continuous lines show the beginnings and 
ends of 10 hours long nights. 
The filled and open circles denote mid eclipse
epochs occurring inside and outside such nights.
The $T_{\mathrm{A1}}=  10$ hour time intervals of eclipses 
are denoted with thick continuous or thin dashed lines.
The tilted open and closed triangles show the
$T_{\mathrm{A2}}=7$ and  $T_{\mathrm{A3}}=3$ hour limits. }
\label{Eclipses}
\end{figure}

{\val Algol}'s observable night time mid eclipse epochs 
occur in groups of three
separated with a period of 19 days and 
we also discovered this period in CC\cite{Jet13}.
This phenomenon is displayed in Fig. \ref{Eclipses}.
First, a mid eclipse epoch occurs in the end of the night.
After three days, the next one occurs close to midnight.
After another three days, a mid eclipse epoch
occurs in the beginning of the night.
Then, the next observable night-time mid eclipse epoch
occurs after 13 days.
Naked eye observations could easily lead to
the discovery of this $3+3+13$ days regularity. 
One could speculate that this is one of the reasons,
why the prime number 13 is still considered unlucky.
This would be consistent with our result that,
if the brightest phases of {\val Algol} were considered lucky
then the eclipses (i.e. the dimmer phases) were considered unlucky.
The 2.85 days period is exactly equal to 57/20 days.
This means that after $57 = 3 \times 19$ days
the eclipses returned exactly 
to the same moment of the night (see Fig. \ref{Eclipses}).
All $D=1$ days in CC have a prognosis combination ``GGG'',
while all $D=20$ days have ``SSS''.
Perhaps this regular separation of 19 days was also inspired by {\val Algol}.

Only a skilled naked eye observer would have been able to discover
the minor exceptions from the $3+3+13$ days regularity.
{\val Algol}'s eclipses last $T_{A1}=10$ hours.
Naked eye can detect brightness 
differences of $0.^{\mathrm{m}}1$ in ideal observing conditions. 
Hence, an eclipse detection is theoretically {\val possible} 
for $T_{A2}=7$ hours when {\val Algol} is more than
$0.^{\mathrm{m}}1$ dimmer than 
its brightest suitable comparison star $\gamma$ And
(Fig. \ref{Eclipses}: tilted open triangle limits).
This detection could become {\val certain} for $T_{A3}=3$ hours
when {\val Algol} is also at least $0.^{\mathrm{m}}1$ dimmer than all
its other suitable comparison stars 
$\zeta$ Per, $\epsilon$ Per, $\gamma$ Per, $\delta$ Per and $\beta$ Tri
(Fig. \ref{Eclipses}: tilted closed triangle limits).
During the 57 days eclipse repetition cycle,
only two mid eclipse epochs outside the 10 hour night time limits
would qualify as {\val certain} observable eclipses 
(Fig. \ref{Eclipses}: open circles at 19th and 48th days).
However, a {\val certain} detection of these two events
would have been very difficult so close to dawn and dusk.
The same argument is true for three additional {\val possible} eclipse
detections (Fig. \ref{Eclipses}: open circles at 11th, 31st and 54th days).

Here, our statistical analysis of SWs 
giving the largest impact on the $P_{\mathrm{A}}$ signal 
reveals that {\val Algol} was represented as {\val Horus}.
The lucky prognoses were most likely connected to {\val Algol}'s brightest phase. 
{\val Sakhmet} may have represented {\val Algol} after eclipses,
and {\val Wedjat} during periods close to the Full {\val Moon}.
To the Ancient Egyptians, {\val Algol}'s cycle may have symbolised the familiar 
events of LE1 and LE2.
At Aa, {\val Re} sends the Eye of Horus ({\val Wedjat})
to destroy the rebels, as in LE2.
At Ab, {\val Horus}  enters the {\it ``foreign land''} in $g_{\mathrm{i}}(7,9)$, 
where he {\it ``smote him who rebelled''}, as in LE1 or LE2.
The {\it ``will is written''} for him in $g_{\mathrm{i}}(28,3)$ at
the beginning of an eclipse --
the only {\bf g}$_{\mathrm{i}}$ vector of {\val Horus} overlapping
the thick line centered at Ac in Fig \ref{Horus}a.
After an eclipse, {\val Wedjat} returns as {\val Sakhmet} who is pacified 
immediately after Ad, as in LE2. And a new cycle begins.

{\val Followers} and {\val Ennead} may have represented {\val Pleiades}.
Thus, these two, 
together with
{\val Horus},
{\val Re},
{\val Wedjat} and
{\val Sakhmet},
give the largest impact on the $P_{\mathrm{A}}$ signal.

The two periods, $P_{\mathrm{A}}$ and $P_{\mathrm{M}}$,
regulate the assignment of mythological texts to specific days of the year.
The {\val Moon} strongly regulates the times described as lucky 
for {\val Heaven} and for {\val Earth}
(Figs \ref{Earth}c and \ref{Heaven}c).
The unlucky prognoses of 
 {\val \nameref{Sethpara}}
are clearly associated
with the phases of the {\val Moon} (Fig \ref{Seth}d).
Other SWs follow $P_{\mathrm{A}}$ or $P_{\mathrm{M}}$,
like 
{\val Busiris},  
{\val Heart}, 
{\val Osiris} and 
{\val Man}
(Figs 
\ref{Busiris}, 
\ref{Heart},  
\ref{Osiris} and 
\ref{Man}).
We show no figures for
{\val Heliopolis},
{\val Enemy},
{\val Rebel},
{\val Thoth},
{\val Onnophris},
{\val Nut},
{\val Nun},
{\val Abydos},
{\val Lion}
and
{\val Flame}
which also reach $Q_{\mathrm{z}} \le 0.2$ with $P_{\mathrm{A}}$ or $P_{\mathrm{M}}$.
All these regularities can not simply be dismissed as a coincidence,
let alone with the possible errors of
$\sigma_{\mathrm{t}} \approx \pm 0.5$ or $\pm 1.5$ days.

\section*{Conclusions}

What was the origin of the phenomenon that occurred every third day, 
but always 3~hours and 36~minutes earlier than before, 
and caught the attention of Ancient Egyptians?
Our statistical analysis leads us to argue that
the mythological texts of CC contain astrophysical information about 
{\val Algol}.
In 1596, Fabricius discovered the first variable star, {\val Mira}.
Holwarda determined its eleven month period 44 years later. 
In 1669, Montanari discovered the second variable star, {\val Algol}.
Goodricke\cite{Goo83} determined the 2.867 days period of {\val Algol} in 1783.
All these astronomical discoveries were made with naked eye.
Since then, they have become milestones of natural sciences.
Our statistical analysis of CC confirms that all these milestones
should be shifted about three millennia backwards in time.

\section*{Acknowledgments}
We thank L. Alha, T. Hackman, T. Lind\'{e}n, K. Muinonen and H. Oja 
for their comments on the manuscript.
This work has made use of 
NASA's Astrophysics Data System (ADS) services.


{\bf Data Availability:}
The authors confirm that all data underlying
these findings are fully available without restriction.
Tables 2--4 and the Python 3.0 program {\val tableS1.py}
have been deposited to Dryad
(http://dx.doi.org/10.5061/dryad.tj4qg).

\clearpage

\section*{Supporting Information}

\subsection*{S1 Fig}
\label{Fragment}
{\bf Text of Cairo Calendar page rto VIII.}
Inside our superimposed rectangle is the hieratic writing 
for the word {\val Horus}.
Reprinted from  Leitz\cite{Lei94}
under a CC BY license, with permission from 
Harrassowitz Verlag,
original copyright [1994].

\begin{figure}[h]
\resizebox{12.5cm}{!}{\includegraphics{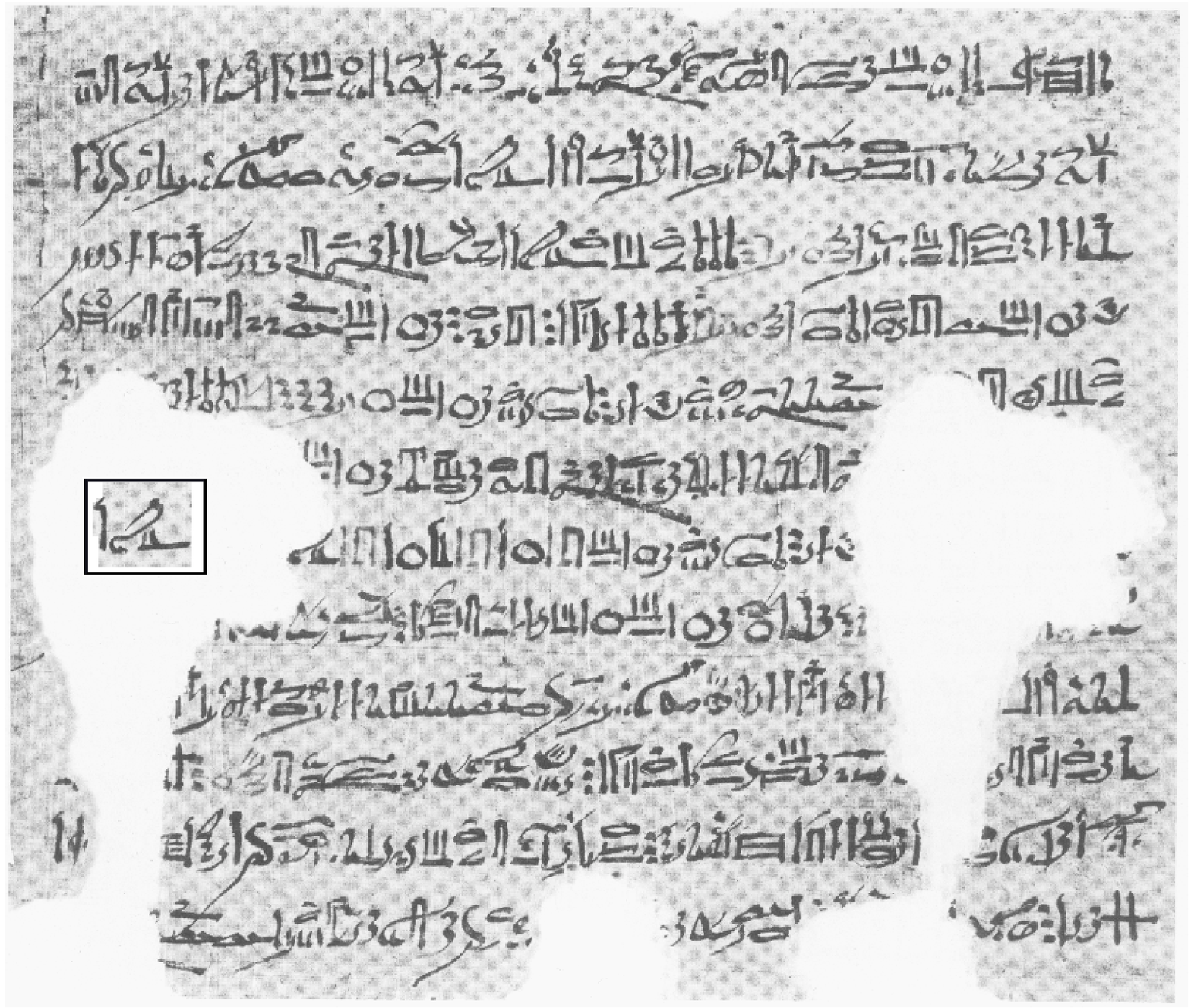}} 
\end{figure}

\clearpage

\subsection*{S1 Table}
\label{S1table}
{\bf Analysis results for all SWs}
Day ($D$), month ($M$) of
lucky ($g_{\mathrm{i}}$) and unlucky ($s_{\mathrm{i}}$) time points,
their phase ($\phi_{\mathrm{i}}$),
phase angle ($\Theta_{\mathrm{i}}$),
direction of their {\bf R} vector ($\Theta_{\mathrm{R}}$)
and differences
$\Delta \Theta_{\mathrm{i}}=\Theta_{\mathrm{i}}-\Theta_{\mathrm{R}}$
with Eq (11) for $P_{\mathrm{A}}=2.85$ days
and Eq (12)  for $P_{\mathrm{M}}=29.6$ days.
The binomial distribution parameters are
 $n_1$, $n_2$, $q_B$ for $Q_{\mathrm{B}}$.
Note that the parameters are given in the order of increasing
$\Delta \Theta_{\mathrm{i}}$, $n_1$ and $n_2$.
All values mentioned in text are marked in bold.
We also make available the code of a Python 3.0 program {\it tableS1.py}
which can be downloaded on Dryad
(http://dx.doi.org/10.5061/dryad.tj4qg).
This program can be used to reproduce and replicate all analysis
results given in S1 Table.

\renewcommand{\arraystretch}{0.82}
\begin{adjustwidth}{-2.25in}{0in}
\begin{tabular}{rrrrrrrrrrrrrrrrrrr}
\multicolumn{19}{c}{ ~  } \\
\multicolumn{19}{c}{Horus:$~n_G={\bf 14}$,$~N_G={\bf 177}$,$~q_B=0.0791$} \\ 
\multicolumn{9}{c}{$P=   2.85 $, $\Theta_R=  11   $}
& ~~ &
\multicolumn{9}{c}{$P=  29.60 $, $\Theta_R= -75   $}
  \\ 
\multicolumn{9}{c}{Periodicity:~$z=3.6$,~$Q_z={\bf 0.03}$,~$z_x={\bf 3.5}$}
& ~~ &
\multicolumn{9}{c}{No~periodicity}
  \\ 
\cline{1-9} \cline{11-19}
 D & M &$g_i   $&$\phi_i$&$\Theta_i$&$\Delta \Theta_i$&$n_1$&$n_2$&$Q_B$& &
 D & M &$g_i   $&$\phi_i$&$\Theta_i$&$\Delta \Theta_i$&$n_1$&$n_2$&$Q_B$\\
\cline{1-9} \cline{11-19}
19&12& 348.33&  0.04&{\bf  13}&  1& 1&  4&0.2808& &
27& 3&  86.33& -0.20&{\bf -73}&  2& 1&  2&0.1519 \\
14& 2&  43.33&  0.02&{\bf  6}&  5& 2&  5&0.0532& &
27& 1&  26.33& -0.23&{\bf -82}&  8& 2&  7&0.1006 \\
27& 1&  26.33&  0.05&{\bf  19}&  7& 3&  9&0.0289& &
28& 3&  87.33& -0.17& -60& 14& 3& 16&0.1279 \\
\end{tabular}
            
 ~   \\     
\begin{tabular}{rrrrrrrrrrrrrrrrrrr}
\multicolumn{19}{c}{ ~  } \\
\multicolumn{19}{c}{Re:$~n_G={\bf 31}$,$~N_G=177$,$~q_B=0.1751$} \\ 
\multicolumn{9}{c}{$P=   2.85 $, $\Theta_R=  23   $}
& ~~ &
\multicolumn{9}{c}{$P=  29.60 $, $\Theta_R=  -7   $}
  \\ 
\multicolumn{9}{c}{Periodicity:~$z=2.9$,~$Q_z={\bf 0.05}$,~$z_x={\bf 2.5}$}
& ~~ &
\multicolumn{9}{c}{No~periodicity}
  \\ 
\cline{1-9} \cline{11-19}
 D & M &$g_i   $&$\phi_i$&$\Theta_i$&$\Delta \Theta_i$&$n_1$&$n_2$&$Q_B$& &
 D & M &$g_i   $&$\phi_i$&$\Theta_i$&$\Delta \Theta_i$&$n_1$&$n_2$&$Q_B$\\
\cline{1-9} \cline{11-19}
17& 2&  46.33&  0.07&  25&  2& 1&  1&0.1751& &
 3& 2&  32.33& -0.03&  -9&  3& 1&  6&0.6850 \\
24& 3&  83.33&  0.05&  19&  4& 2&  8&0.4217& &
 2& 6& 151.33& -0.01&  -2&  5& 2&  9&0.4854 \\
25&10& 294.33&  0.09&  32&  8& 3& 16&0.5496& &
 1& 6& 150.33& -0.04& -14&  7& 3& 14&0.4549 \\
\end{tabular}
            
 ~   \\     
\begin{tabular}{rrrrrrrrrrrrrrrrrrr}
\multicolumn{19}{c}{ ~  } \\
\multicolumn{19}{c}{Wedjat:$~n_G=4$,$~N_G=177$,$~q_B=0.0226$} \\ 
\multicolumn{9}{c}{$P=   2.85 $, $\Theta_R=  -8   $}
& ~~ &
\multicolumn{9}{c}{$P=  29.60 $, $\Theta_R=   6   $}
  \\ 
\multicolumn{9}{c}{Weak~periodicity:~$z=2.1$,~$Q_z=0.1$,~$z_x={\bf 2.0}$}
& ~~ &
\multicolumn{9}{c}{Periodicity:~$z=2.9$,~$Q_z=0.05$,~$z_x={\bf 2.9}$}
  \\ 
\cline{1-9} \cline{11-19}
 D & M &$g_i   $&$\phi_i$&$\Theta_i$&$\Delta \Theta_i$&$n_1$&$n_2$&$Q_B$& &
 D & M &$g_i   $&$\phi_i$&$\Theta_i$&$\Delta \Theta_i$&$n_1$&$n_2$&$Q_B$\\
\cline{1-9} \cline{11-19}
 1& 5& 120.33&  0.04&  13& 21& 1& 29&0.4846& &
30&10& 299.33& -0.01&  -2&  8& 1& 12&0.2399 \\
 6& 9& 245.33& -0.11& -38& 30& 2& 43&0.2537& &
 3& 2&  32.33& -0.03&  -9& 15& 2& 20&0.0742 \\
30&10& 299.33& -0.16& -57& 49& 3& 73&0.2285& &
 1& 5& 120.33& -0.05& -19& 25& 3& 36&0.0474 \\
 3& 2&  32.33&  0.16&  57& 65& 4& 84&0.1228& &
 6& 9& 245.33&  0.17&  61& 55& 4& 72&0.0804 \\
\cline{1-9} \cline{11-19}
\multicolumn{19}{c}{ ~  } \\
\multicolumn{19}{c}{Wedjat:$~n_S=1$,$~N_S=105$,$~q_B=0.0095$} \\ 
\multicolumn{9}{c}{$P=   2.85 $, $\Theta_R=-101   $}
& ~~ &
\multicolumn{9}{c}{$P=  29.60 $, $\Theta_R=  95   $}
  \\ 
\multicolumn{9}{c}{No~periodicity}
& ~~ &
\multicolumn{9}{c}{No~periodicity}
   \\ 
\cline{1-9} \cline{11-19}
 D & M &$s_i   $&$\phi_i$&$\Theta_i$&$               $&$   $&$   $&$   $& &
 D & M &$s_i   $&$\phi_i$&$\Theta_i$&$               $&$   $&$   $&$   $\\
\cline{1-9} \cline{11-19}
10& 6& 159.33& -0.28&-101&   &  &   &      & &
10& 6& 159.33&  0.26&  95&   &  &   &       \\
\cline{1-9} \cline{11-19}
\end{tabular}
            
 ~   \\     
\begin{tabular}{rrrrrrrrrrrrrrrrrrr}
\multicolumn{19}{c}{ ~  } \\
\multicolumn{19}{c}{Son:$~n_G=6$,$~N_G=177$,$~q_B=0.0339$} \\ 
\multicolumn{9}{c}{$P=   2.85 $, $\Theta_R=  18   $}
& ~~ &
\multicolumn{9}{c}{$P=  29.60 $, $\Theta_R= -82   $}
  \\ 
\multicolumn{9}{c}{Weak~periodicity:~$z=2.3$,~$Q_z=0.1$,~$z_x={\bf 2.0}$}
& ~~ &
\multicolumn{9}{c}{No~periodicity}
  \\ 
\cline{1-9} \cline{11-19}
 D & M &$g_i   $&$\phi_i$&$\Theta_i$&$\Delta \Theta_i$&$n_1$&$n_2$&$Q_B$& &
 D & M &$g_i   $&$\phi_i$&$\Theta_i$&$\Delta \Theta_i$&$n_1$&$n_2$&$Q_B$\\
\cline{1-9} \cline{11-19}
24& 3&  83.33&  0.05&  19&  1& 1&  2&0.0666& &
28& 3&  87.33& -0.17& -60& 21& 1& 22&0.5317 \\
 1& 5& 120.33&  0.04&  13&  6& 2&  6&0.0157& &
24& 3&  83.33& -0.30&-109& 27& 2& 27&0.2325 \\
21& 3&  80.33&  0.00&   0& 18& 3& 27&0.0622& &
 1& 5& 120.33& -0.05& -19& 63& 3& 59&0.3233 \\
\end{tabular}
            
 ~   \\     
\begin{tabular}{rrrrrrrrrrrrrrrrrrr}
\multicolumn{19}{c}{ ~  } \\
\multicolumn{19}{c}{Sakhmet:$~n_G={\bf 3}$,$~N_G=177$,$~q_B=0.0169$} \\ 
\multicolumn{9}{c}{$P=   2.85 $, $\Theta_R= -33   $}
& ~~ &
\multicolumn{9}{c}{$P=  29.60 $, $\Theta_R=   1   $}
  \\ 
\multicolumn{9}{c}{Weak~periodicity:~$z=2.2$,~$Q_z={\bf 0.1}$,~$z_x={\bf 1.6}$}
& ~~ &
\multicolumn{9}{c}{No~periodicity}
  \\ 
\cline{1-9} \cline{11-19}
 D & M &$g_i   $&$\phi_i$&$\Theta_i$&$\Delta \Theta_i$&$n_1$&$n_2$&$Q_B$& &
 D & M &$g_i   $&$\phi_i$&$\Theta_i$&$\Delta \Theta_i$&$n_1$&$n_2$&$Q_B$\\
\cline{1-9} \cline{11-19}
29& 5& 148.33& -0.14& -51& 18& 1& 25&0.3478& &
 1& 5& 120.33& -0.05& -19& 20& 1& 29&0.3909 \\
 9& 5& 128.33& -0.16& -57& 24& 2& 33&0.1075& &
29& 5& 148.33& -0.11& -39& 39& 2& 56&0.2454 \\
 1& 5& 120.33&  0.04&  13& 45& 3& 56&0.0696& &
 9& 5& 128.33&  0.22&  78& 77& 3& 97&0.2273 \\
\cline{1-9} \cline{11-19}
\multicolumn{19}{c}{ ~  } \\
\multicolumn{19}{c}{Sakhmet:$~n_S={\bf 3}$,$~N_S={\bf 105}$,$~q_B=0.0286$} \\ 
\multicolumn{9}{c}{$P=   2.85 $, $\Theta_R= -86   $}
& ~~ &
\multicolumn{9}{c}{$P=  29.60 $, $\Theta_R=  78   $}
  \\ 
\multicolumn{9}{c}{Periodicity:~$z=3.0$,~$Q_z={\bf 0.05}$,~$z_x=0.0$}
& ~~ &
\multicolumn{9}{c}{No~periodicity}
   \\ 
\cline{1-9} \cline{11-19}
 D & M &$s_i   $&$\phi_i$&$\Theta_i$&$\Delta \Theta_i$&$n_1$&$n_2$&$Q_B$& &
 D & M &$s_i   $&$\phi_i$&$\Theta_i$&$\Delta \Theta_i$&$n_1$&$n_2$&$Q_B$\\
\cline{1-9} \cline{11-19}
13& 6& 162.33& -0.23&{\bf -82}&  4& 1&  3&0.0833& &
 7&10& 276.33&  0.22&  78&  0& 1&  1&0.0286 \\
 7&10& 276.33& -0.23&{\bf -82}&  4& 2&  4&0.0047& &
13& 6& 162.33&  0.37& 132& 54& 2& 32&0.2323 \\
27& 8& 236.33& -0.26&{\bf -95}&  8&{\bf 3}&{\bf  6}&{\bf 0.0004}& &
27& 8& 236.33& -0.13& -48&126& 3& 70&0.3233 \\
\cline{1-9} \cline{11-19}
\end{tabular}
            
 ~   \\     
\begin{tabular}{rrrrrrrrrrrrrrrrrrr}
\multicolumn{19}{c}{ ~  } \\
\multicolumn{19}{c}{Followers:$~n_G={\bf 15}$,$~N_G=177$,$~q_B=0.0847$} \\ 
\multicolumn{9}{c}{$P=   2.85 $, $\Theta_R=  22   $}
& ~~ &
\multicolumn{9}{c}{$P=  29.60 $, $\Theta_R=  57   $}
  \\ 
\multicolumn{9}{c}{Weak~periodicity:~$z=1.6$,~$Q_z={\bf 0.2}$,~$z_x={\bf 1.4}$}
& ~~ &
\multicolumn{9}{c}{No~periodicity}
  \\ 
\cline{1-9} \cline{11-19}
 D & M &$g_i   $&$\phi_i$&$\Theta_i$&$\Delta \Theta_i$&$n_1$&$n_2$&$Q_B$& &
 D & M &$g_i   $&$\phi_i$&$\Theta_i$&$\Delta \Theta_i$&$n_1$&$n_2$&$Q_B$\\
\cline{1-9} \cline{11-19}
21& 5& 140.33&  0.05&  19&  3& 1&  3&0.2333& &
 6& 9& 245.33&  0.17&  61&  4& 1&  5&0.3577 \\
 1& 5& 120.33&  0.04&  13&  9& 2& 12&0.2705& &
10& 2&  39.33&  0.21&  76& 18& 2& 20&0.5148 \\
29&11& 328.33&  0.02&   6& 15& 3& 25&0.3567& &
 3&10& 272.33&  0.08&  30& 28& 3& 30&0.4731 \\
\end{tabular}
            
 ~   \\     
\begin{tabular}{rrrrrrrrrrrrrrrrrrr}
\multicolumn{19}{c}{ ~  } \\
\multicolumn{19}{c}{Ennead:$~n_G=18$,$~N_G={\bf 177}$,$~q_B=0.1017$} \\ 
\multicolumn{9}{c}{$P=   2.85 $, $\Theta_R=  46   $}
& ~~ &
\multicolumn{9}{c}{$P=  29.60 $, $\Theta_R=  64   $}
  \\ 
\multicolumn{9}{c}{Weak~periodicity:~$z=2.2$,~$Q_z=0.1$,~$z_x={\bf 1.1}$}
& ~~ &
\multicolumn{9}{c}{No~periodicity}
  \\ 
\cline{1-9} \cline{11-19}
 D & M &$g_i   $&$\phi_i$&$\Theta_i$&$\Delta \Theta_i$&$n_1$&$n_2$&$Q_B$& &
 D & M &$g_i   $&$\phi_i$&$\Theta_i$&$\Delta \Theta_i$&$n_1$&$n_2$&$Q_B$\\
\cline{1-9} \cline{11-19}
10& 3&  69.33&  0.14&  51&  4& 1&  3&0.2751& &
 9& 2&  38.33&  0.18&  64&  0& 1&  1&0.1017 \\
24& 5& 143.33&  0.11&  38&  9& 2&  9&0.2310& &
10& 3&  69.33&  0.22&  81& 17& 2& 19&0.5893 \\
18& 9& 257.33&  0.11&  38&  9& 3& 10&0.0731& &
 8& 8& 217.33&  0.22&  81& 17& 3& 20&0.3327 \\
\end{tabular}
            
 ~   \\     
\begin{tabular}{rrrrrrrrrrrrrrrrrrr}
\multicolumn{19}{c}{ ~  } \\
\multicolumn{19}{c}{Heliopolis:$~n_G=4$,$~N_G=177$,$~q_B=0.0226$} \\ 
\multicolumn{9}{c}{$P=   2.85 $, $\Theta_R= -70   $}
& ~~ &
\multicolumn{9}{c}{$P=  29.60 $, $\Theta_R= 138   $}
  \\ 
\multicolumn{9}{c}{Weak~periodicity:~$z=2.1$,~$Q_z=0.1$,~$z_x={\bf 0.2}$}
& ~~ &
\multicolumn{9}{c}{No~periodicity}
  \\ 
\cline{1-9} \cline{11-19}
 D & M &$g_i   $&$\phi_i$&$\Theta_i$&$\Delta \Theta_i$&$n_1$&$n_2$&$Q_B$& &
 D & M &$g_i   $&$\phi_i$&$\Theta_i$&$\Delta \Theta_i$&$n_1$&$n_2$&$Q_B$\\
\cline{1-9} \cline{11-19}
26& 3&  85.33& -0.25& -88& 18& 1& 14&0.2739& &
 9& 7& 188.33&  0.24&  88& 50& 1& 44&0.6342 \\
 9& 7& 188.33& -0.11& -38& 32& 2& 34&0.1789& &
10& 2&  39.33&  0.21&  76& 62& 2& 53&0.3374 \\
19& 8& 228.33& -0.07& -25& 45& 3& 47&0.0900& &
19& 8& 228.33& -0.40&-146& 76& 3& 67&0.1933 \\
10& 2&  39.33& -0.39&-139& 69& 4& 69&0.0711& &
26& 3&  85.33& -0.24& -85&137& 4&121&0.2928 \\
\cline{1-9} \cline{11-19}
\multicolumn{19}{c}{ ~  } \\
\multicolumn{19}{c}{Heliopolis:$~n_S=4$,$~N_S=105$,$~q_B=0.0381$} \\ 
\multicolumn{9}{c}{$P=   2.85 $, $\Theta_R=-168   $}
& ~~ &
\multicolumn{9}{c}{$P=  29.60 $, $\Theta_R= 132   $}
  \\ 
\multicolumn{9}{c}{Weak~periodicity:~$z=2.3$,~$Q_z=0.1$,~$z_x=-2.2$}
& ~~ &
\multicolumn{9}{c}{No~periodicity}
   \\ 
\cline{1-9} \cline{11-19}
 D & M &$s_i   $&$\phi_i$&$\Theta_i$&$\Delta \Theta_i$&$n_1$&$n_2$&$Q_B$& &
 D & M &$s_i   $&$\phi_i$&$\Theta_i$&$\Delta \Theta_i$&$n_1$&$n_2$&$Q_B$\\
\cline{1-9} \cline{11-19}
22&11& 321.33& -0.44&-158& 10& 1&  8&0.2671& &
10& 6& 159.33&  0.26&  95& 37& 1& 22&0.5745 \\
16&11& 315.33&  0.46& 164& 27& 2& 18&0.1487& &
16&11& 315.33& -0.47&-167& 61& 2& 41&0.4662 \\
 3&12& 332.33&  0.42& 152& 40& 3& 27&0.0820& &
 3&12& 332.33&  0.11&  39& 93& 3& 64&0.4424 \\
10& 6& 159.33& -0.28&-101& 67& 4& 45&0.0915& &
22&11& 321.33& -0.26& -95&134& 4& 85&0.4070 \\
\cline{1-9} \cline{11-19}
\end{tabular}
            
 ~   \\     
\begin{tabular}{rrrrrrrrrrrrrrrrrrr}
\multicolumn{19}{c}{ ~  } \\
\multicolumn{19}{c}{Enemy:$~n_G=6$,$~N_G=177$,$~q_B=0.0339$} \\ 
\multicolumn{9}{c}{$P=   2.85 $, $\Theta_R= 140   $}
& ~~ &
\multicolumn{9}{c}{$P=  29.60 $, $\Theta_R=  77   $}
  \\ 
\multicolumn{9}{c}{Weak~periodicity:~$z=1.7$,~$Q_z=0.2$,~$z_x={\bf -1.0}$}
& ~~ &
\multicolumn{9}{c}{No~periodicity}
  \\ 
\cline{1-9} \cline{11-19}
 D & M &$g_i   $&$\phi_i$&$\Theta_i$&$\Delta \Theta_i$&$n_1$&$n_2$&$Q_B$& &
 D & M &$g_i   $&$\phi_i$&$\Theta_i$&$\Delta \Theta_i$&$n_1$&$n_2$&$Q_B$\\
\cline{1-9} \cline{11-19}
23&10& 292.33&  0.39& 139&  1& 1&  2&0.0666& &
 9& 4&  98.33&  0.20&  73&  3& 1&  4&0.1289 \\
 9& 4&  98.33&  0.32& 114& 27& 2& 19&0.1345& &
 9& 2&  38.33&  0.18&  64& 13& 2& 16&0.1007 \\
 9& 2&  38.33&  0.26&  95& 46& 3& 36&0.1219& &
13& 2&  42.33&  0.31& 112& 35& 3& 36&0.1219 \\
\end{tabular}
            
 ~   \\     
\begin{tabular}{rrrrrrrrrrrrrrrrrrr}
\multicolumn{19}{c}{ ~  } \\
\multicolumn{19}{c}{Earth:$~n_G={\bf 19}$,$~N_G=177$,$~q_B=0.1073$} \\ 
\multicolumn{9}{c}{$P=   2.85 $, $\Theta_R= -61   $}
& ~~ &
\multicolumn{9}{c}{$P=  29.60 $, $\Theta_R= -27   $}
  \\ 
\multicolumn{9}{c}{No~periodicity}
& ~~ &
\multicolumn{9}{c}{Periodicity:~$z=6.6$,~$Q_z={\bf 0.001}$,~$z_x={\bf 5.3}$}
  \\ 
\cline{1-9} \cline{11-19}
 D & M &$g_i   $&$\phi_i$&$\Theta_i$&$\Delta \Theta_i$&$n_1$&$n_2$&$Q_B$& &
 D & M &$g_i   $&$\phi_i$&$\Theta_i$&$\Delta \Theta_i$&$n_1$&$n_2$&$Q_B$\\
\cline{1-9} \cline{11-19}
 2& 4&  91.33& -0.14& -51& 10& 1&  9&0.6401& &
 1& 1&   0.33& -0.11& -39& 12& 1& 17&0.8549 \\
29& 5& 148.33& -0.14& -51& 10& 2& 10&0.2924& &
29& 5& 148.33& -0.11& -39& 12& 2& 18&0.5902 \\
26& 3&  85.33& -0.25& -88& 28& 3& 25&0.5119& &
 1& 6& 150.33& -0.04& -14& 13& 3& 19&0.3344 \\
\end{tabular}
            
 ~   \\     
\begin{tabular}{rrrrrrrrrrrrrrrrrrr}
\multicolumn{19}{c}{ ~  } \\
\multicolumn{19}{c}{Heaven:$~n_G={\bf 19}$,$~N_G={\bf 177}$,$~q_B=0.1073$} \\ 
\multicolumn{9}{c}{$P=   2.85 $, $\Theta_R=  10   $}
& ~~ &
\multicolumn{9}{c}{$P=  29.60 $, $\Theta_R=   0   $}
  \\ 
\multicolumn{9}{c}{No~periodicity}
& ~~ &
\multicolumn{9}{c}{Periodicity:~$z=3.4$,~$Q_z={\bf 0.03}$,~$z_x={\bf 3.4}$}
  \\ 
\cline{1-9} \cline{11-19}
 D & M &$g_i   $&$\phi_i$&$\Theta_i$&$\Delta \Theta_i$&$n_1$&$n_2$&$Q_B$& &
 D & M &$g_i   $&$\phi_i$&$\Theta_i$&$\Delta \Theta_i$&$n_1$&$n_2$&$Q_B$\\
\cline{1-9} \cline{11-19}
 2&12& 331.33&  0.07&  25& 15& 1& 22&0.9178& &
 1& 8& 210.33& -0.01&  -5&  4& 1&  7&0.5484 \\
 1& 3&  60.33& -0.02&  -6& 16& 2& 23&0.7236& &
 1& 7& 180.33& -0.03&  -9&  9& 2& 16&0.5247 \\
 1& 7& 180.33&  0.09&  32& 22& 3& 29&0.6153& &
 5& 1&   4.33&  0.03&  10& 10& 3& 18&0.3036 \\
\end{tabular}
            
 ~   \\     
\begin{tabular}{rrrrrrrrrrrrrrrrrrr}
\multicolumn{19}{c}{ ~  } \\
\multicolumn{19}{c}{Busiris:$~n_G={\bf 4}$,$~N_G=177$,$~q_B=0.0226$} \\ 
\multicolumn{9}{c}{$P=   2.85 $, $\Theta_R=  19   $}
& ~~ &
\multicolumn{9}{c}{$P=  29.60 $, $\Theta_R=   3   $}
  \\ 
\multicolumn{9}{c}{No~periodicity}
& ~~ &
\multicolumn{9}{c}{Periodicity:~$z=3.0$,~$Q_z={\bf 0.05}$,~$z_x={\bf 3.0}$}
  \\ 
\cline{1-9} \cline{11-19}
 D & M &$g_i   $&$\phi_i$&$\Theta_i$&$\Delta \Theta_i$&$n_1$&$n_2$&$Q_B$& &
 D & M &$g_i   $&$\phi_i$&$\Theta_i$&$\Delta \Theta_i$&$n_1$&$n_2$&$Q_B$\\
\cline{1-9} \cline{11-19}
 2& 8& 211.33& -0.04& -13& 32& 1& 43&0.6258& &
 2& 8& 211.33&  0.02&   8&  5& 1&  6&0.1282 \\
28&10& 297.33&  0.14&  51& 32& 2& 48&0.2957& &
30& 7& 209.33& -0.05& -17& 20& 2& 29&0.1391 \\
 6& 7& 185.33& -0.16& -57& 76& 3& 93&0.3511& &
28&10& 297.33& -0.07& -26& 29& 3& 45&0.0813 \\
30& 7& 209.33&  0.26&  95& 76& 4& 97&0.1773& &
 6& 7& 185.33&  0.14&  51& 48& 4& 63&0.0543 \\
\cline{1-9} \cline{11-19}
\multicolumn{19}{c}{ ~  } \\
\multicolumn{19}{c}{Busiris:$~n_S=3$,$~N_S=105$,$~q_B=0.0286$} \\ 
\multicolumn{9}{c}{$P=   2.85 $, $\Theta_R=-123   $}
& ~~ &
\multicolumn{9}{c}{$P=  29.60 $, $\Theta_R=-111   $}
  \\ 
\multicolumn{9}{c}{No~periodicity}
& ~~ &
\multicolumn{9}{c}{No~periodicity}
   \\ 
\cline{1-9} \cline{11-19}
 D & M &$s_i   $&$\phi_i$&$\Theta_i$&$\Delta \Theta_i$&$n_1$&$n_2$&$Q_B$& &
 D & M &$s_i   $&$\phi_i$&$\Theta_i$&$\Delta \Theta_i$&$n_1$&$n_2$&$Q_B$\\
\cline{1-9} \cline{11-19}
14& 5& 133.33& -0.40&-145& 22& 1& 19&0.4235& &
26& 2&  55.33& -0.25& -90& 21& 1& 16&0.3711 \\
26& 5& 145.33& -0.19& -69& 53& 2& 39&0.3068& &
26& 5& 145.33& -0.21& -75& 35& 2& 27&0.1798 \\
26& 2&  55.33&  0.23&  82&155& 3& 87&0.4543& &
14& 5& 133.33&  0.39& 139&110& 3& 68&0.3076 \\
\cline{1-9} \cline{11-19}
\end{tabular}
            
 ~   \\     
\begin{tabular}{rrrrrrrrrrrrrrrrrrr}
\multicolumn{19}{c}{ ~  } \\
\multicolumn{19}{c}{Rebel:$~n_G=3$,$~N_G=177$,$~q_B=0.0169$} \\ 
\multicolumn{9}{c}{$P=   2.85 $, $\Theta_R= 125   $}
& ~~ &
\multicolumn{9}{c}{$P=  29.60 $, $\Theta_R=  11   $}
  \\ 
\multicolumn{9}{c}{No~periodicity}
& ~~ &
\multicolumn{9}{c}{Weak~periodicity:~$z=1.6$,~$Q_z=0.2$,~$z_x={\bf 1.6}$}
  \\ 
\cline{1-9} \cline{11-19}
 D & M &$g_i   $&$\phi_i$&$\Theta_i$&$\Delta \Theta_i$&$n_1$&$n_2$&$Q_B$& &
 D & M &$g_i   $&$\phi_i$&$\Theta_i$&$\Delta \Theta_i$&$n_1$&$n_2$&$Q_B$\\
\cline{1-9} \cline{11-19}
 2& 1&   1.33&  0.28& 101& 24& 1& 16&0.2393& &
 1& 8& 210.33& -0.01&  -5& 16& 1& 20&0.2896 \\
 7& 9& 246.33&  0.25&  88& 36& 2& 30&0.0915& &
 2& 1&   1.33& -0.07& -26& 38& 2& 49&0.2017 \\
 1& 8& 210.33& -0.39&-139& 96& 3& 87&0.1837& &
 7& 9& 246.33&  0.20&  73& 62& 3& 81&0.1587 \\
\cline{1-9} \cline{11-19}
\multicolumn{19}{c}{ ~  } \\
\multicolumn{19}{c}{Rebel:$~n_S=3$,$~N_S=105$,$~q_B=0.0286$} \\ 
\multicolumn{9}{c}{$P=   2.85 $, $\Theta_R= 107   $}
& ~~ &
\multicolumn{9}{c}{$P=  29.60 $, $\Theta_R=-162   $}
  \\ 
\multicolumn{9}{c}{No~periodicity}
& ~~ &
\multicolumn{9}{c}{No~periodicity}
   \\ 
\cline{1-9} \cline{11-19}
 D & M &$s_i   $&$\phi_i$&$\Theta_i$&$\Delta \Theta_i$&$n_1$&$n_2$&$Q_B$& &
 D & M &$s_i   $&$\phi_i$&$\Theta_i$&$\Delta \Theta_i$&$n_1$&$n_2$&$Q_B$\\
\cline{1-9} \cline{11-19}
12& 2&  41.33&  0.32& 114&  7& 1&  5&0.1349& &
20& 2&  49.33& -0.45&-163&  0& 1&  1&0.0286 \\
23&12& 352.33&  0.44& 158& 51& 2& 38&0.2962& &
23&12& 352.33& -0.22& -77& 85& 2& 63&0.5406 \\
20& 2&  49.33&  0.12&  44& 63& 3& 42&0.1180& &
12& 2&  41.33&  0.28& 100& 98& 3& 69&0.3155 \\
\cline{1-9} \cline{11-19}
\end{tabular}
            
 ~   \\     
\begin{tabular}{rrrrrrrrrrrrrrrrrrr}
\multicolumn{19}{c}{ ~  } \\
\multicolumn{19}{c}{Thoth:$~n_G=10$,$~N_G=177$,$~q_B=0.0565$} \\ 
\multicolumn{9}{c}{$P=   2.85 $, $\Theta_R= 115   $}
& ~~ &
\multicolumn{9}{c}{$P=  29.60 $, $\Theta_R= -39   $}
  \\ 
\multicolumn{9}{c}{No~periodicity}
& ~~ &
\multicolumn{9}{c}{Weak~periodicity:~$z=2.1$,~$Q_z=0.1$,~$z_x={\bf 1.3}$}
  \\ 
\cline{1-9} \cline{11-19}
 D & M &$g_i   $&$\phi_i$&$\Theta_i$&$\Delta \Theta_i$&$n_1$&$n_2$&$Q_B$& &
 D & M &$g_i   $&$\phi_i$&$\Theta_i$&$\Delta \Theta_i$&$n_1$&$n_2$&$Q_B$\\
\cline{1-9} \cline{11-19}
 9& 4&  98.33&  0.32& 114&  1& 1&  1&0.0565& &
29& 5& 148.33& -0.11& -39&  1& 1&  2&0.1098 \\
13& 7& 192.33&  0.30& 107&  8& 2&  6&0.0411& &
28& 5& 147.33& -0.14& -51& 11& 2& 15&0.2066 \\
26& 4& 115.33&  0.28& 101& 14& 3&  9&0.0117& &
 1& 4&  90.33& -0.07& -24& 15& 3& 21&0.1124 \\
\end{tabular}
            
 ~   \\     
\begin{tabular}{rrrrrrrrrrrrrrrrrrr}
\multicolumn{19}{c}{ ~  } \\
\multicolumn{19}{c}{Onnophris:$~n_G=7$,$~N_G=177$,$~q_B=0.0395$} \\ 
\multicolumn{9}{c}{$P=   2.85 $, $\Theta_R=-138   $}
& ~~ &
\multicolumn{9}{c}{$P=  29.60 $, $\Theta_R= -47   $}
  \\ 
\multicolumn{9}{c}{No~periodicity}
& ~~ &
\multicolumn{9}{c}{Weak~periodicity:~$z=2.1$,~$Q_z=0.1$,~$z_x={\bf 1.0}$}
  \\ 
\cline{1-9} \cline{11-19}
 D & M &$g_i   $&$\phi_i$&$\Theta_i$&$\Delta \Theta_i$&$n_1$&$n_2$&$Q_B$& &
 D & M &$g_i   $&$\phi_i$&$\Theta_i$&$\Delta \Theta_i$&$n_1$&$n_2$&$Q_B$\\
\cline{1-9} \cline{11-19}
28& 7& 207.33& -0.44&-158& 20& 1& 14&0.4316& &
28& 6& 177.33& -0.13& -46&  1& 1&  2&0.0775 \\
16& 2&  45.33& -0.28&-101& 37& 2& 24&0.2451& &
28& 7& 207.33& -0.11& -41&  6& 2&  7&0.0288 \\
 1&12& 330.33& -0.28&-101& 37& 3& 27&0.0894& &
28& 3&  87.33& -0.17& -60& 14& 3& 16&0.0235 \\
\end{tabular}
            
 ~   \\     
\begin{tabular}{rrrrrrrrrrrrrrrrrrr}
\multicolumn{19}{c}{ ~  } \\
\multicolumn{19}{c}{Nut:$~n_G=3$,$~N_G=177$,$~q_B=0.0169$} \\ 
\multicolumn{9}{c}{$P=   2.85 $, $\Theta_R=  55   $}
& ~~ &
\multicolumn{9}{c}{$P=  29.60 $, $\Theta_R=-105   $}
  \\ 
\multicolumn{9}{c}{No~periodicity}
& ~~ &
\multicolumn{9}{c}{Weak~periodicity:~$z=1.8$,~$Q_z=0.2$,~$z_x={\bf -0.1}$}
  \\ 
\cline{1-9} \cline{11-19}
 D & M &$g_i   $&$\phi_i$&$\Theta_i$&$\Delta \Theta_i$&$n_1$&$n_2$&$Q_B$& &
 D & M &$g_i   $&$\phi_i$&$\Theta_i$&$\Delta \Theta_i$&$n_1$&$n_2$&$Q_B$\\
\cline{1-9} \cline{11-19}
22&12& 351.33&  0.09&  32& 24& 1& 28&0.3804& &
22&12& 351.33& -0.25& -90& 16& 1& 13&0.1993 \\
18& 7& 197.33&  0.05&  19& 37& 2& 44&0.1711& &
28& 1&  27.33& -0.19& -70& 35& 2& 26&0.0714 \\
28& 1&  27.33&  0.40& 145& 90& 3&100&0.2407& &
18& 7& 197.33& -0.45&-163& 57& 3& 45&0.0408 \\
\cline{1-9} \cline{11-19}
\multicolumn{19}{c}{ ~  } \\
\multicolumn{19}{c}{Nut:$~n_S=4$,$~N_S=105$,$~q_B=0.0381$} \\ 
\multicolumn{9}{c}{$P=   2.85 $, $\Theta_R=  81   $}
& ~~ &
\multicolumn{9}{c}{$P=  29.60 $, $\Theta_R= 139   $}
  \\ 
\multicolumn{9}{c}{No~periodicity}
& ~~ &
\multicolumn{9}{c}{No~periodicity}
   \\ 
\cline{1-9} \cline{11-19}
 D & M &$s_i   $&$\phi_i$&$\Theta_i$&$\Delta \Theta_i$&$n_1$&$n_2$&$Q_B$& &
 D & M &$s_i   $&$\phi_i$&$\Theta_i$&$\Delta \Theta_i$&$n_1$&$n_2$&$Q_B$\\
\cline{1-9} \cline{11-19}
17& 8& 226.33&  0.23&  82&  1& 1&  3&0.1100& &
12& 2&  41.33&  0.28& 100& 39& 1& 24&0.6063 \\
12& 2&  41.33&  0.32& 114& 33& 2& 24&0.2321& &
17& 8& 226.33& -0.47&-170& 51& 2& 34&0.3735 \\
20& 2&  49.33&  0.12&  44& 37& 3& 27&0.0820& &
20& 2&  49.33& -0.45&-163& 58& 3& 41&0.2046 \\
 4&10& 273.33& -0.28&-101&178& 4&105&0.5703& &
 4&10& 273.33&  0.12&  42& 98& 4& 66&0.2434 \\
\cline{1-9} \cline{11-19}
\end{tabular}
            
 ~   \\     
\begin{tabular}{rrrrrrrrrrrrrrrrrrr}
\multicolumn{19}{c}{ ~  } \\
\multicolumn{19}{c}{Heart:$~n_G=27$,$~N_G=177$,$~q_B=0.1525$} \\ 
\multicolumn{9}{c}{$P=   2.85 $, $\Theta_R= -91   $}
& ~~ &
\multicolumn{9}{c}{$P=  29.60 $, $\Theta_R=  47   $}
  \\ 
\multicolumn{9}{c}{No~periodicity}
& ~~ &
\multicolumn{9}{c}{No~periodicity}
  \\ 
\cline{1-9} \cline{11-19}
 D & M &$g_i   $&$\phi_i$&$\Theta_i$&$\Delta \Theta_i$&$n_1$&$n_2$&$Q_B$& &
 D & M &$g_i   $&$\phi_i$&$\Theta_i$&$\Delta \Theta_i$&$n_1$&$n_2$&$Q_B$\\
\cline{1-9} \cline{11-19}
 9& 1&   8.33& -0.26& -95&  4& 1&  4&0.4842& &
 9& 1&   8.33&  0.16&  59& 11& 1&  8&0.7340 \\
 6& 3&  65.33& -0.26& -95&  4& 2&  5&0.1695& &
 6& 3&  65.33&  0.09&  32& 15& 2& 14&0.6531 \\
16& 2&  45.33& -0.28&-101& 10& 3& 11&0.2289& &
 9& 2&  38.33&  0.18&  64& 16& 3& 16&0.4502 \\
\end{tabular}
            
 ~   \\     
\begin{tabular}{rrrrrrrrrrrrrrrrrrr}
\multicolumn{19}{c}{ ~  } \\
\multicolumn{19}{c}{Nun:$~n_G=7$,$~N_G=177$,$~q_B=0.0395$} \\ 
\multicolumn{9}{c}{$P=   2.85 $, $\Theta_R= -38   $}
& ~~ &
\multicolumn{9}{c}{$P=  29.60 $, $\Theta_R= -52   $}
  \\ 
\multicolumn{9}{c}{No~periodicity}
& ~~ &
\multicolumn{9}{c}{No~periodicity}
  \\ 
\cline{1-9} \cline{11-19}
 D & M &$g_i   $&$\phi_i$&$\Theta_i$&$\Delta \Theta_i$&$n_1$&$n_2$&$Q_B$& &
 D & M &$g_i   $&$\phi_i$&$\Theta_i$&$\Delta \Theta_i$&$n_1$&$n_2$&$Q_B$\\
\cline{1-9} \cline{11-19}
 1& 1&   0.33& -0.07& -25& 13& 1& 14&0.4316& &
30& 2&  59.33& -0.11& -41& 11& 1& 10&0.3320 \\
12& 7& 191.33& -0.05& -19& 19& 2& 25&0.2599& &
 1& 1&   0.33& -0.11& -39& 13& 2& 14&0.1039 \\
19& 2&  48.33& -0.23& -82& 44& 3& 55&0.3715& &
30& 5& 149.33& -0.07& -26& 25& 3& 30&0.1140 \\
\end{tabular}
            
 ~   \\     
\begin{tabular}{rrrrrrrrrrrrrrrrrrr}
\multicolumn{19}{c}{ ~  } \\
\multicolumn{19}{c}{Seth:$~n_G=5$,$~N_G=177$,$~q_B=0.0282$} \\ 
\multicolumn{9}{c}{$P=   2.85 $, $\Theta_R=  11   $}
& ~~ &
\multicolumn{9}{c}{$P=  29.60 $, $\Theta_R= -73   $}
  \\ 
\multicolumn{9}{c}{No~periodicity}
& ~~ &
\multicolumn{9}{c}{No~periodicity}
  \\ 
\cline{1-9} \cline{11-19}
 D & M &$g_i   $&$\phi_i$&$\Theta_i$&$\Delta \Theta_i$&$n_1$&$n_2$&$Q_B$& &
 D & M &$g_i   $&$\phi_i$&$\Theta_i$&$\Delta \Theta_i$&$n_1$&$n_2$&$Q_B$\\
\cline{1-9} \cline{11-19}
27& 1&  26.33&  0.05&  19&  8& 1&  9&0.2273& &
27& 3&  86.33& -0.20& -73&  0& 1&  1&0.0282 \\
27& 3&  86.33&  0.11&  38& 27& 2& 40&0.3126& &
27& 1&  26.33& -0.23& -82& 10& 2&  9&0.0252 \\
13&12& 342.33& -0.07& -25& 36& 3& 55&0.2031& &
29& 3&  88.33& -0.13& -48& 24& 3& 24&0.0293 \\
\end{tabular}
            
 ~   \\     
\begin{tabular}{rrrrrrrrrrrrrrrrrrr}
\multicolumn{19}{c}{ ~  } \\
\multicolumn{19}{c}{Osiris:$~n_G=8$,$~N_G=177$,$~q_B=0.0452$} \\ 
\multicolumn{9}{c}{$P=   2.85 $, $\Theta_R= -82   $}
& ~~ &
\multicolumn{9}{c}{$P=  29.60 $, $\Theta_R=  93   $}
  \\ 
\multicolumn{9}{c}{No~periodicity}
& ~~ &
\multicolumn{9}{c}{No~periodicity}
  \\ 
\cline{1-9} \cline{11-19}
 D & M &$g_i   $&$\phi_i$&$\Theta_i$&$\Delta \Theta_i$&$n_1$&$n_2$&$Q_B$& &
 D & M &$g_i   $&$\phi_i$&$\Theta_i$&$\Delta \Theta_i$&$n_1$&$n_2$&$Q_B$\\
\cline{1-9} \cline{11-19}
16& 2&  45.33& -0.28&-101& 19& 1& 18&0.5650& &
11& 4& 100.33&  0.27&  98&  4& 1&  3&0.1296 \\
 6& 7& 185.33& -0.16& -57& 25& 2& 23&0.2791& &
 6& 7& 185.33&  0.14&  51& 42& 2& 43&0.5846 \\
 1&10& 270.33& -0.33&-120& 38& 3& 39&0.2579& &
16& 2&  45.33&  0.41& 149& 56& 3& 50&0.3949 \\
\end{tabular}
            
 ~   \\     
\begin{tabular}{rrrrrrrrrrrrrrrrrrr}
\multicolumn{19}{c}{ ~  } \\
\multicolumn{19}{c}{Abydos:$~n_G=3$,$~N_G=177$,$~q_B=0.0169$} \\ 
\multicolumn{9}{c}{$P=   2.85 $, $\Theta_R= -14   $}
& ~~ &
\multicolumn{9}{c}{$P=  29.60 $, $\Theta_R= 151   $}
  \\ 
\multicolumn{9}{c}{No~periodicity}
& ~~ &
\multicolumn{9}{c}{No~periodicity}
  \\ 
\cline{1-9} \cline{11-19}
 D & M &$g_i   $&$\phi_i$&$\Theta_i$&$\Delta \Theta_i$&$n_1$&$n_2$&$Q_B$& &
 D & M &$g_i   $&$\phi_i$&$\Theta_i$&$\Delta \Theta_i$&$n_1$&$n_2$&$Q_B$\\
\cline{1-9} \cline{11-19}
18& 5& 137.33&  0.00&   0& 14& 1& 22&0.3135& &
18& 5& 137.33& -0.48&-172& 36& 1& 29&0.3909 \\
11& 4& 100.33&  0.02&   6& 20& 2& 30&0.0915& &
11& 4& 100.33&  0.27&  98& 54& 2& 43&0.1650 \\
28& 7& 207.33& -0.44&-158&144& 3&146&0.4507& &
28& 7& 207.33& -0.11& -41&168& 3&159&0.5066 \\
\cline{1-9} \cline{11-19}
\multicolumn{19}{c}{ ~  } \\
\multicolumn{19}{c}{Abydos:$~n_S=2$,$~N_S=105$,$~q_B=0.0190$} \\ 
\multicolumn{9}{c}{$P=   2.85 $, $\Theta_R=   3   $}
& ~~ &
\multicolumn{9}{c}{$P=  29.60 $, $\Theta_R= 129   $}
  \\ 
\multicolumn{9}{c}{No~periodicity}
& ~~ &
\multicolumn{9}{c}{Weak~periodicity:~$z=1.9$,~$Q_z=0.1$,~$z_x=-0.8$}
   \\ 
\cline{1-9} \cline{11-19}
 D & M &$s_i   $&$\phi_i$&$\Theta_i$&$\Delta \Theta_i$&$n_1$&$n_2$&$Q_B$& &
 D & M &$s_i   $&$\phi_i$&$\Theta_i$&$\Delta \Theta_i$&$n_1$&$n_2$&$Q_B$\\
\cline{1-9} \cline{11-19}
13& 3&  72.33&  0.19&  69& 66& 1& 26&0.3935& &
13& 3&  72.33&  0.33& 117& 12& 1&  9&0.1589 \\
13& 8& 222.33& -0.18& -63& 66& 2& 28&0.0990& &
13& 8& 222.33&  0.39& 141& 12& 2& 12&0.0211 \\
\cline{1-9} \cline{11-19}
\end{tabular}
            
 ~   \\     
\begin{tabular}{rrrrrrrrrrrrrrrrrrr}
\multicolumn{19}{c}{ ~  } \\
\multicolumn{19}{c}{Lion:$~n_G=1$,$~N_G=177$,$~q_B=0.0056$} \\ 
\multicolumn{9}{c}{$P=   2.85 $, $\Theta_R=-177   $}
& ~~ &
\multicolumn{9}{c}{$P=  29.60 $, $\Theta_R= -24   $}
  \\ 
\multicolumn{9}{c}{No~periodicity}
& ~~ &
\multicolumn{9}{c}{No~periodicity}
  \\ 
\cline{1-9} \cline{11-19}
 D & M &$g_i   $&$\phi_i$&$\Theta_i$&$               $&$   $&$   $&$   $& &
 D & M &$g_i   $&$\phi_i$&$\Theta_i$&$               $&$   $&$   $&$   $\\
\cline{1-9} \cline{11-19}
 1& 4&  90.33& -0.49&-177&   &  &   &      & &
 1& 4&  90.33& -0.07& -24&   &  &   &       \\
\cline{1-9} \cline{11-19}
\multicolumn{19}{c}{ ~  } \\
\multicolumn{19}{c}{Lion:$~n_S=6$,$~N_S=105$,$~q_B=0.0571$} \\ 
\multicolumn{9}{c}{$P=   2.85 $, $\Theta_R=-162   $}
& ~~ &
\multicolumn{9}{c}{$P=  29.60 $, $\Theta_R=-100   $}
  \\ 
\multicolumn{9}{c}{No~periodicity}
& ~~ &
\multicolumn{9}{c}{Weak~periodicity:~$z=2.2$,~$Q_z=0.1$,~$z_x=-0.1$}
   \\ 
\cline{1-9} \cline{11-19}
 D & M &$s_i   $&$\phi_i$&$\Theta_i$&$\Delta \Theta_i$&$n_1$&$n_2$&$Q_B$& &
 D & M &$s_i   $&$\phi_i$&$\Theta_i$&$\Delta \Theta_i$&$n_1$&$n_2$&$Q_B$\\
\cline{1-9} \cline{11-19}
24& 2&  53.33& -0.47&-171&  8& 1&  6&0.2975& &
25& 2&  54.33& -0.28&-102&  2& 1&  2&0.1110 \\
18&10& 287.33& -0.37&-133& 29& 2& 19&0.2966& &
24& 2&  53.33& -0.32&-114& 14& 2&  9&0.0899 \\
26&10& 295.33&  0.44& 158& 40& 3& 28&0.2135& &
19&10& 288.33& -0.38&-136& 36& 3& 25&0.1692 \\
 3& 8& 212.33&  0.32& 114& 84& 4& 56&0.3987& &
18&10& 287.33& -0.41&-148& 48& 4& 33&0.1172 \\
25& 2&  54.33& -0.12& -44&118& 5& 79&0.4735& &
26&10& 295.33& -0.14& -51& 49& 5& 34&0.0425 \\
19&10& 288.33& -0.02&  -6&156& 6&101&0.5207& &
 3& 8& 212.33&  0.06&  20&119& 6& 70&0.2102 \\
\cline{1-9} \cline{11-19}
\end{tabular}
            
 ~   \\     
\begin{tabular}{rrrrrrrrrrrrrrrrrrr}
\multicolumn{19}{c}{ ~  } \\
\multicolumn{19}{c}{Man:$~n_G=5$,$~N_G=177$,$~q_B=0.0282$} \\ 
\multicolumn{9}{c}{$P=   2.85 $, $\Theta_R= -27   $}
& ~~ &
\multicolumn{9}{c}{$P=  29.60 $, $\Theta_R=   3   $}
  \\ 
\multicolumn{9}{c}{No~periodicity}
& ~~ &
\multicolumn{9}{c}{No~periodicity}
  \\ 
\cline{1-9} \cline{11-19}
 D & M &$g_i   $&$\phi_i$&$\Theta_i$&$\Delta \Theta_i$&$n_1$&$n_2$&$Q_B$& &
 D & M &$g_i   $&$\phi_i$&$\Theta_i$&$\Delta \Theta_i$&$n_1$&$n_2$&$Q_B$\\
\cline{1-9} \cline{11-19}
16&10& 285.33& -0.07& -25&  2& 1&  5&0.1335& &
 3& 5& 122.33&  0.01&   5&  2& 1&  3&0.0824 \\
 6& 5& 125.33& -0.21& -76& 49& 2& 60&0.5083& &
 4& 5& 123.33&  0.05&  17& 14& 2& 22&0.1272 \\
 4& 5& 123.33&  0.09&  32& 59& 3& 73&0.3404& &
 6& 5& 125.33&  0.12&  42& 39& 3& 53&0.1886 \\
\end{tabular}
            
 ~   \\     
\begin{tabular}{rrrrrrrrrrrrrrrrrrr}
\multicolumn{19}{c}{Flame:$~n_S={\bf 4}$,$~N_S={\bf 105}$,$~q_B=0.0381$} \\ 
\multicolumn{9}{c}{$P=   2.85 $, $\Theta_R= 123   $}
& ~~ &
\multicolumn{9}{c}{$P=  29.60 $, $\Theta_R=  85   $}
  \\ 
\multicolumn{9}{c}{No~periodicity}
& ~~ &
\multicolumn{9}{c}{Periodicity:~$z=3.6$,~$Q_z={\bf 0.03}$,~$z_x=0.0$}
   \\ 
\cline{1-9} \cline{11-19}
 D & M &$s_i   $&$\phi_i$&$\Theta_i$&$\Delta \Theta_i$&$n_1$&$n_2$&$Q_B$& &
 D & M &$s_i   $&$\phi_i$&$\Theta_i$&$\Delta \Theta_i$&$n_1$&$n_2$&$Q_B$\\
\cline{1-9} \cline{11-19}
10& 7& 189.33&  0.25&  88& 35& 1& 19&0.5219& &
11& 1&  10.33&  0.23&  83&  2& 1&  1&0.0381 \\
11& 1&  10.33&  0.44& 158& 35& 2& 20&0.1759& &
10& 7& 189.33&  0.28& 100& 15& 2& 10&0.0533 \\
11& 5& 130.33& -0.46&-164& 73& 3& 47&0.2657& &
11& 5& 130.33&  0.28& 103& 17& 3& 12&0.0094 \\
 7& 5& 126.33&  0.14&  51& 73& 4& 50&0.1224& &
 7& 5& 126.33&  0.15&  54& 31&{\bf 4}&{\bf 17}&{\bf 0.0034} \\
\cline{1-9} \cline{11-19}
\end{tabular}
            
 ~   \\     
\begin{tabular}{rrrrrrrrrrrrrrrrrrr}
\multicolumn{19}{c}{ ~  } \\
\multicolumn{19}{c}{Eye:$~n_G=4$,$~N_G=177$,$~q_B=0.0226$} \\ 
\multicolumn{9}{c}{$P=   2.85 $, $\Theta_R= -24   $}
& ~~ &
\multicolumn{9}{c}{$P=  29.60 $, $\Theta_R= 115   $}
  \\ 
\multicolumn{9}{c}{No~periodicity}
& ~~ &
\multicolumn{9}{c}{No~periodicity}
  \\ 
\cline{1-9} \cline{11-19}
 D & M &$g_i   $&$\phi_i$&$\Theta_i$&$\Delta \Theta_i$&$n_1$&$n_2$&$Q_B$& &
 D & M &$g_i   $&$\phi_i$&$\Theta_i$&$\Delta \Theta_i$&$n_1$&$n_2$&$Q_B$\\
\cline{1-9} \cline{11-19}
10&12& 339.33& -0.12& -44& 21& 1& 32&0.5188& &
10&12& 339.33&  0.35& 124& 10& 1& 11&0.2223 \\
19&12& 348.33&  0.04&  13& 36& 2& 51&0.3208& &
10& 4&  99.33&  0.24&  85& 29& 2& 24&0.1016 \\
 8& 8& 217.33&  0.07&  25& 49& 3& 63&0.1706& &
 8& 8& 217.33&  0.22&  81& 34& 3& 27&0.0225 \\
10& 4&  99.33& -0.33&-120& 96& 4&111&0.2430& &
19&12& 348.33& -0.35&-126&119& 4&110&0.2381 \\
\cline{1-9} \cline{11-19}
\multicolumn{19}{c}{ ~  } \\
\multicolumn{19}{c}{Eye:$~n_S=5$,$~N_S=105$,$~q_B=0.0476$} \\ 
\multicolumn{9}{c}{$P=   2.85 $, $\Theta_R=  81   $}
& ~~ &
\multicolumn{9}{c}{$P=  29.60 $, $\Theta_R= 124   $}
  \\ 
\multicolumn{9}{c}{No~periodicity}
& ~~ &
\multicolumn{9}{c}{No~periodicity}
   \\ 
\cline{1-9} \cline{11-19}
 D & M &$s_i   $&$\phi_i$&$\Theta_i$&$\Delta \Theta_i$&$n_1$&$n_2$&$Q_B$& &
 D & M &$s_i   $&$\phi_i$&$\Theta_i$&$\Delta \Theta_i$&$n_1$&$n_2$&$Q_B$\\
\cline{1-9} \cline{11-19}
 7& 7& 186.33&  0.19&  69& 11& 1& 11&0.4153& &
11&11& 310.33&  0.37& 132&  8& 1&  8&0.3232 \\
20& 4& 109.33&  0.18&  63& 18& 2& 14&0.1414& &
14&11& 313.33&  0.47& 168& 45& 2& 26&0.3531 \\
 3& 8& 212.33&  0.32& 114& 33& 3& 25&0.1141& &
 7& 7& 186.33&  0.18&  64& 60& 3& 36&0.2446 \\
14&11& 313.33& -0.25& -88&169& 4& 96&0.6759& &
20& 4& 109.33& -0.42&-153& 84& 4& 55&0.2660 \\
11&11& 310.33& -0.30&-107&172& 5&101&0.5291& &
 3& 8& 212.33&  0.06&  20&104& 5& 68&0.2228 \\
\cline{1-9} \cline{11-19}
\end{tabular}
            
 ~   \\     
\begin{tabular}{rrrrrrrrrrrrrrrrrrr}
\multicolumn{19}{c}{ ~  } \\
\multicolumn{19}{c}{Fire:$~n_G=4$,$~N_G=177$,$~q_B=0.0226$} \\ 
\multicolumn{9}{c}{$P=   2.85 $, $\Theta_R= 122   $}
& ~~ &
\multicolumn{9}{c}{$P=  29.60 $, $\Theta_R=   4   $}
  \\ 
\multicolumn{9}{c}{No~periodicity}
& ~~ &
\multicolumn{9}{c}{No~periodicity}
  \\ 
\cline{1-9} \cline{11-19}
 D & M &$g_i   $&$\phi_i$&$\Theta_i$&$\Delta \Theta_i$&$n_1$&$n_2$&$Q_B$& &
 D & M &$g_i   $&$\phi_i$&$\Theta_i$&$\Delta \Theta_i$&$n_1$&$n_2$&$Q_B$\\
\cline{1-9} \cline{11-19}
19& 1&  18.33&  0.25&  88& 33& 1& 25&0.4353& &
29& 8& 238.33& -0.07& -24& 28& 1& 40&0.5992 \\
29& 8& 238.33&  0.44& 158& 36& 2& 31&0.1548& &
30& 5& 149.33& -0.07& -26& 30& 2& 44&0.2621 \\
30& 5& 149.33&  0.21&  76& 46& 3& 38&0.0542& &
 7& 8& 216.33&  0.19&  68& 65& 3& 81&0.2770 \\
\end{tabular}
            
 ~   \\     
\begin{tabular}{rrrrrrrrrrrrrrrrrrr}
\multicolumn{19}{c}{ ~  } \\
\multicolumn{19}{c}{Majesty:$~n_G=13$,$~N_G=177$,$~q_B=0.0734$} \\ 
\multicolumn{9}{c}{$P=   2.85 $, $\Theta_R= -64   $}
& ~~ &
\multicolumn{9}{c}{$P=  29.60 $, $\Theta_R=  10   $}
  \\ 
\multicolumn{9}{c}{No~periodicity}
& ~~ &
\multicolumn{9}{c}{No~periodicity}
  \\ 
\cline{1-9} \cline{11-19}
 D & M &$g_i   $&$\phi_i$&$\Theta_i$&$\Delta \Theta_i$&$n_1$&$n_2$&$Q_B$& &
 D & M &$g_i   $&$\phi_i$&$\Theta_i$&$\Delta \Theta_i$&$n_1$&$n_2$&$Q_B$\\
\cline{1-9} \cline{11-19}
 2& 2&  31.33& -0.19& -69&  6& 1&  2&0.1415& &
 2& 8& 211.33&  0.02&   8&  3& 1&  4&0.2630 \\
18& 1&  17.33& -0.11& -38& 26& 2& 24&0.5348& &
 5& 4&  94.33&  0.07&  25& 15& 2& 17&0.3582 \\
 9& 1&   8.33& -0.26& -95& 31& 3& 28&0.3391& &
 2& 2&  31.33& -0.06& -22& 32& 3& 44&0.6364 \\
\end{tabular}
            
 ~   \\     
\begin{tabular}{rrrrrrrrrrrrrrrrrrr}
\multicolumn{19}{c}{ ~  } \\
\multicolumn{19}{c}{Shu:$~n_G=3$,$~N_G=177$,$~q_B=0.0169$} \\ 
\multicolumn{9}{c}{$P=   2.85 $, $\Theta_R=   5   $}
& ~~ &
\multicolumn{9}{c}{$P=  29.60 $, $\Theta_R=-158   $}
  \\ 
\multicolumn{9}{c}{No~periodicity}
& ~~ &
\multicolumn{9}{c}{No~periodicity}
  \\ 
\cline{1-9} \cline{11-19}
 D & M &$g_i   $&$\phi_i$&$\Theta_i$&$\Delta \Theta_i$&$n_1$&$n_2$&$Q_B$& &
 D & M &$g_i   $&$\phi_i$&$\Theta_i$&$\Delta \Theta_i$&$n_1$&$n_2$&$Q_B$\\
\cline{1-9} \cline{11-19}
21& 3&  80.33&  0.00&   0&  5& 1&  6&0.0975& &
21& 3&  80.33& -0.40&-146& 13& 1&  7&0.1128 \\
30&10& 299.33& -0.16& -57& 62& 2& 84&0.4176& &
16& 5& 135.33&  0.45& 163& 39& 2& 25&0.0666 \\
16& 5& 135.33&  0.30& 107&102& 3&121&0.3372& &
30&10& 299.33& -0.01&  -2&156& 3&151&0.4725 \\
\cline{1-9} \cline{11-19}
\multicolumn{19}{c}{ ~  } \\
\multicolumn{19}{c}{Shu:$~n_S=1$,$~N_S=105$,$~q_B=0.0095$} \\ 
\multicolumn{9}{c}{$P=   2.85 $, $\Theta_R=  13   $}
& ~~ &
\multicolumn{9}{c}{$P=  29.60 $, $\Theta_R= -99   $}
  \\ 
\multicolumn{9}{c}{No~periodicity}
& ~~ &
\multicolumn{9}{c}{No~periodicity}
   \\ 
\cline{1-9} \cline{11-19}
 D & M &$s_i   $&$\phi_i$&$\Theta_i$&$               $&$   $&$   $&$   $& &
 D & M &$s_i   $&$\phi_i$&$\Theta_i$&$               $&$   $&$   $&$   $\\
\cline{1-9} \cline{11-19}
22&10& 291.33&  0.04&  13&   &  &   &      & &
22&10& 291.33& -0.28& -99&   &  &   &       \\
\cline{1-9} \cline{11-19}
\end{tabular}
            
 ~   \\     
\begin{tabular}{rrrrrrrrrrrrrrrrrrr}
\multicolumn{19}{c}{ ~  } \\
\multicolumn{19}{c}{Sobek:$~n_G=1$,$~N_G=177$,$~q_B=0.0056$} \\ 
\multicolumn{9}{c}{$P=   2.85 $, $\Theta_R=  25   $}
& ~~ &
\multicolumn{9}{c}{$P=  29.60 $, $\Theta_R= 107   $}
  \\ 
\multicolumn{9}{c}{No~periodicity}
& ~~ &
\multicolumn{9}{c}{No~periodicity}
  \\ 
\cline{1-9} \cline{11-19}
 D & M &$g_i   $&$\phi_i$&$\Theta_i$&$               $&$   $&$   $&$   $& &
 D & M &$g_i   $&$\phi_i$&$\Theta_i$&$               $&$   $&$   $&$   $\\
\cline{1-9} \cline{11-19}
11& 6& 160.33&  0.07&  25&   &  &   &      & &
11& 6& 160.33&  0.30& 107&   &  &   &       \\
\cline{1-9} \cline{11-19}
\multicolumn{19}{c}{ ~  } \\
\multicolumn{19}{c}{Sobek:$~n_S=3$,$~N_S=105$,$~q_B=0.0286$} \\ 
\multicolumn{9}{c}{$P=   2.85 $, $\Theta_R= 161   $}
& ~~ &
\multicolumn{9}{c}{$P=  29.60 $, $\Theta_R= -89   $}
  \\ 
\multicolumn{9}{c}{No~periodicity}
& ~~ &
\multicolumn{9}{c}{No~periodicity}
   \\ 
\cline{1-9} \cline{11-19}
 D & M &$s_i   $&$\phi_i$&$\Theta_i$&$\Delta \Theta_i$&$n_1$&$n_2$&$Q_B$& &
 D & M &$s_i   $&$\phi_i$&$\Theta_i$&$\Delta \Theta_i$&$n_1$&$n_2$&$Q_B$\\
\cline{1-9} \cline{11-19}
26&10& 295.33&  0.44& 158&  3& 1&  2&0.0563& &
25& 8& 234.33& -0.20& -73& 17& 1& 13&0.3140 \\
17& 1&  16.33& -0.46&-164& 35& 2& 20&0.1105& &
26&10& 295.33& -0.14& -51& 38& 2& 24&0.1492 \\
25& 8& 234.33&  0.04&  13&148& 3& 91&0.4838& &
17& 1&  16.33&  0.43& 156&115& 3& 67&0.2998 \\
\cline{1-9} \cline{11-19}
\end{tabular}
\end{adjustwidth}

\end{document}